\documentclass[11pt,a4paper]{article}
\pdfoutput=1
\usepackage{jheppub}
\usepackage{graphicx}
\usepackage{epsfig}  
\usepackage{epsf}    
\usepackage{dcolumn}
\usepackage{bm}
\usepackage{dcolumn}
\usepackage{textcomp}
\usepackage{float}
\usepackage{subfig}

\usepackage[]{hyperref}
  \hypersetup{
  bookmarks=true,         
  unicode=true,         
  pdftoolbar=true,     
  pdfmenubar=true,     
  pdffitwindow=true,     
  pdfstartview={FitH},    
  pdfsubject={Sterilenus@DUNE},  
  pdfnewwindow=true,     
  pdfcreator={RevTeX},
  colorlinks=true,     
  linkcolor=red,   
  citecolor=blue,    
  filecolor=black,  
  urlcolor=blue,           
  }
\usepackage{hypcap}

\def\nue{{\nu_e}}

\def\numu{{\nu_{\mu}}}

\def\anu{{\bar\nu}}
\newcommand{\eg}{{\it e.g.}}
\newcommand{\ie}{{\it i.e.}}
\newcommand{\etc}{{\it etc.}}

\newcommand{\beq}{\begin{equation}}
\newcommand{\eeq}{\end{equation}}
\newcommand{\beqa}{\begin{eqnarray}}
\newcommand{\eeqa}{\end{eqnarray}}

\newcommand{\tx}{{\theta_{12}}}
\newcommand{\ty}{{\theta_{13}}}
\newcommand{\tz}{{\theta_{23}}}
\newcommand{\ta}{{\theta_{14}}}
\newcommand{\tb}{{\theta_{24}}}
\newcommand{\tc}{{\theta_{34}}}

\newcommand{\dcp}{\delta_{\mathrm{CP}}}

\newcommand{\pme}{P_{\mu e}}

\newcommand{\pmebar}{P_{\bar{\mu} \bar{e}}}

\newcommand{\dxx}{\Delta\chi^2}
\newcommand{\dxxmin}{\Delta\chi^2_{\rm{min}}}

\newcommand{\da}{\delta_{13}}
\newcommand{\db}{\delta_{24}}
\newcommand{\dc}{\delta_{34}}

\preprint{\\ FERMILAB-PUB-16-251-T \\ HRI-P-16-07-001}
\title{Capabilities of long-baseline experiments in the presence of a sterile neutrino}

\author[a]{Debajyoti Dutta,}
\author[a]{Raj Gandhi,}
\author[b]{Boris Kayser,} 
\author[a]{Mehedi Masud,} 
{\author[a,c]{Suprabh Prakash$\,$}

\affiliation[a]{Harish-Chandra Research Institute, Chhatnag Road, Jhunsi,
Allahabad 211019, India.}
\affiliation[b]{Theoretical Physics Department, Fermilab, P.O. Box 500, Batavia,
IL 60510 USA.}
\affiliation[c]{School of Physics, Sun Yat-Sen (Zhongshan) University,  Guangzhou 510275, P. R. China.}

\emailAdd{debajyotidutta@hri.res.in}
\emailAdd{raj@hri.res.in}
\emailAdd{boris@fnal.gov}
\emailAdd{masud@hri.res.in}
\emailAdd{prakash3@mail.sysu.edu.cn}

\begin{abstract}
{
Assuming that there is a sterile neutrino, we ask what then is the ability of long-baseline experiments to i)  establish that neutrino oscillation violates CP, ii)  determine the three-neutrino mass ordering, and iii)  determine which CP-violating phase or phases are the cause of any CP violation that may be observed. We find that the ability to establish CP violation and to determine the mass ordering could be very substantial. However, the effects of the sterile neutrino could be quite large, and it might prove very difficult to determine which phase is responsible for an observed CP violation. We explain why a sterile neutrino changes the long-baseline sensitivities to CP violation and to the mass ordering in the ways that it does. We note that long-baseline experiments can probe the presence of sterile neutrinos in a way that is different from, and complementary to, the probes of short-baseline experiments. We explore the question of how large sterile-active mixing angles need to be before long-baseline experiments can detect their effects, or how small they need to be before the interpretation of these experiments can safely disregard the possible existence of sterile neutrinos.
}
\end{abstract}

\keywords{Leptonic CP Violation, Sterile neutrinos, Long-Baseline experiments, DUNE, NO$\nu$A, T2K, T2HK}

\begin{document}
\vspace*{25px}

%====================================================================================
% \author{Debajyoti Dutta}
% \email[Email Address: ]{debajyotidutta@hri.res.in}
% \affiliation{Harish-Chandra Research Institute, Chhatnag Road, Jhunsi, Allahabad 211019, India}
% 
% \author{Raj Gandhi}
% \email[Email Address: ]{raj@hri.res.in}
% \affiliation{Harish-Chandra Research Institute, Chhatnag Road, Jhunsi, Allahabad 211019, India}
% 
% \author{Boris Kayser}
% \email[Email Address: ]{boris@fnal.gov}
% \affiliation{Theoretical Physics Department, Fermilab, P.O. Box 500, Batavia, IL 60510 USA}
% 
% \author{Mehedi Masud}
% \email[Email Address: ]{masud@hri.res.in}
% \affiliation{Harish-Chandra Research Institute, Chhatnag Road, Jhunsi, Allahabad 211019, India}
% 
% \author{Suprabh Prakash\footnote{Corresponding author}}
% \email[Email Address: ]{prakash3@mail.sysu.edu.cn} 
% \affiliation{Harish-Chandra Research Institute, Chhatnag Road, Jhunsi, Allahabad 211019, India}
% \affiliation{School of Physics, Sun Yat-Sen (Zhongshan) University, \\ Guangzhou 510275, P. R. China}

%====================================================================================
\date{July 8, 2016}

%====================================================================================

%====================================================================================

\maketitle
\flushbottom

%====================================================================================
\section{Introduction}
\label{Introduction}
Much of the  activity in neutrino oscillation experiments over the past two decades has focused on the increasingly precise determination of a) the neutrino mass-squared differences, $\delta m^{2}_{ij} = m^{2}_{i} - m^{2}_{j}$,  and b) the mixing angles $\theta_{ij}$ along with the attendant CP phase $\dcp$, which constitute the PMNS \cite{Maki:1962mu, Pontecorvo:1967fh, Gribov:1968kq} mixing matrix $U_{\alpha i}$, with {\it{i, j}} = 1, 2, 3 \& $i
\neq j$ and $\alpha = e,\mu,\tau$. The complex matrix $U$ parameterizes the overlap between the neutrino mass eigenstates $\nu_i$ and the flavour eigenstates $\nu_{\alpha}$. While considerable progress has been achieved towards determining the mixing angles and mass differences to appreciable accuracy, the CP phase $\dcp$ remains almost completely unknown.

 The sources and qualitative nature of the experiments that have helped in this determination have been diverse, with neutrinos from the atmosphere, from reactors, from the sun and from particle accelerators being observed at detectors placed at  baseline lengths spanning a wide range and employing a variety of detection techniques \cite{Ashie:2004mr,Wendell:2010md, Aharmim:2005gt,Ahn:2006zza, Michael:2006rx, Adamson:2007gu,Abe:2013hdq,Abe:2014ugx,Araki:2004mb, Arpesella:2008mt, An:2012eh, Ahn:2012nd, Abe:2012tg}. Global analyses \cite{Fogli:2012ua, Gonzalez-Garcia:2015qrr, Forero:2014bxa} of the data collected  have aided the consolidation of these efforts. This, in turn, has led to the gradual build-up of a consistent three family neutrino paradigm in impressive conformity with what is known, both theoretically and experimentally,  about the lepton and quark sectors of the highly successful Standard Model (SM) of particle physics (when extended, nominally,  to include massive neutrinos). Additionally, these experimental and theoretical efforts have helped formulate  immediate questions that need to be answered by ongoing and planned experiments, which include the determination of a) the presence or absence of CP violation (CPV) in the lepton sector, and b) the mass hierarchy (MH), or ordering, of the neutrino mass eigenstates.

Nonetheless, there are several  ripples in this seemingly smooth fabric, at least some of which could be indicative of underlying new physics (for a discussion see \cite{Altarelli:2014dca}). Such physics, if real, could affect the  interpretation and  sensitivities of present and planned experiments. One such issue relates to signals from a variety of short-baseline experiments \cite{Aguilar:2001ty,
AguilarArevalo:2008rc, Mention:2011rk, Mueller:2011nm, Aguilar-Arevalo:2013pmq}, which hint at the possible existence of short-wavelength oscillations, driven by one or more largely sterile states ({\it{i.e.}} states that do not have  standard weak interactions, but couple indirectly via mixing) with $O$(1 eV$^2$) mass-squared splittings that are significantly larger than the two splittings that characterize the standard 3 family paradigm (referred to in what follows as the 3+0 scenario). These short-wavelength oscillations can have significant effects when the (Baseline $L$)/(Energy $E$) of neutrinos in a beam is $\sim 1$ km/GeV. 

It is generally true, of course,  that for significantly larger values of $L/E$,   the short-wavelength oscillations driven by the large splittings involving an extra neutrino $\nu_4$ will be averaged to an L/E-independent value by the finite energy resolution of a typical detector. However,  even the presence of a single additional sterile  neutrino mass eigenstate (referred to as the 3+1 scenario in what follows) introduces three additional  mixing angles and two additional phases capable of significantly affecting oscillations at large baselines \cite{Klop:2014ima, Gandhi:2015xza}. It has been shown in \cite{Gandhi:2015xza}, using the 1300 km Deep Underground Neutrino Experiment (DUNE) as an example, that if such a neutrino exists in nature,  long-baseline results, interpreted without taking the short-wavelength oscillations into account, could erroneously imply that CP violation is very small or totally absent, when in reality it could be very large. In addition, measurements interpreted as determining the CP-violating phase in the standard 3+0 paradigm could in fact be measuring a linear combination of  one or more phases belonging to the 3+1 sector. These effects, which arise from large interference terms (between the 3+0 and 3+1 sectors) in the appearance probability, are accentuated by the presence of matter, which  brings in contributions from sterile-sector mixings and phases that are quiescent at short baselines. Thus, the presence of the 3+1 (or, more generally, 3 + $n$) sector can significantly blur any conclusions regarding CPV that may be drawn by long-baseline experiments.

Suppose that a $\sim1\textrm{eV}$ mass largely-sterile neutrino does exist. What then is the ability of long-baseline experiments to establish that the leptonic weak interactions violate CP, to determine the mass-ordering of the three established neutrinos, and to determine which CP-violating phase or phases is responsible for any CP violation that may be observed? In this paper, we address these questions. We find that the capacity to establish CP violation and to determine the mass ordering could be very substantial. However, the effects of the extra neutrino could be quite large, and could lead to erroneous or ambiguous conclusions, as already found in the earlier work \cite{Gandhi:2015xza}. We display examples  that  demonstrate the difficulty involved in  determining which phase is causing an observed CP violation. 

This work thus carries forward the approach adopted in \cite{Gandhi:2015xza}. Specifically, we study how expected sensitivities to the MH and CPV at NOvA \cite{Ayres:2007tu}, T2K \cite{Abe:2013hdq}, DUNE \cite{Hewett:2012ns, Adams:2013qkq} and HyperKamiokande (HK) \cite{Abe:2015zbg} are altered in the 3+1 scenario. Some of the other questions we attempt to answer are focused on DUNE as an example. For instance, suppose that the planned program of short-baseline experiments 
\cite{Antonello:2015lea, Zennamo, Gamez, Camilleri}
that will probe the existence of $1\textrm{eV}$-mass sterile neutrinos does not see anything. We ask how tightly one must then bound the sterile-active mixing angles to ensure that DUNE data can be safely interpreted without taking the possible existence of sterile neutrinos into account. 

Early work examining the effects of sterile neutrinos at long-baselines 
considered neutrino-factory experiments with baselines of 3000 km - 
7500 km and muon energies in the range of 20 GeV - 50 GeV, focusing
on effects at both near and far detectors \cite{Donini:2001xp, Dighe:2007uf,
Donini:2008wz, Yasuda:2010rj, Meloni:2010zr}. More recent work 
\cite{Klop:2014ima} includes a study of effects relevant to T2K 
\cite{Abe:2013hdq} and a combined study \cite{Bhattacharya:2011ee}
of T2K, MINOS \cite{Adamson:2014vgd} and reactor experiments
which focuses on the effects of a 3+2 scenario on $\theta_{13}$
measurements. Additionally, \cite{Hollander:2014iha} has analysed the
sensitivity of the experiment now known as DUNE to two sterile neutrinos
whose physics is described by a minimal 3+2 model. Recently, 
in \cite{Berryman:2015nua} the long-baseline effects of one sterile 
neutrino have been studied, and sensitivity calculations in the 
presence of such a neutrino have been presented in 
\cite{Agarwalla:2016mrc, Agarwalla:2016xxa}.

The analyses reported here differ from those referred to above 
in a number of ways, including physics content, approach, questions asked and 
underlying assumptions. While \cite{Hollander:2014iha} considered
a minimal 3+2 model, the illustrative model we consider is 
3+1. \cite{Berryman:2015nua} examines the large allowed
range of $\Delta m^2_{41}$ and focuses on how well DUNE can
constrain the 3+1 and 3+0 hypotheses assuming, in one 
case, that there is no sterile neutrino, and in a second case, 
if there is one sterile neutrino. In the 3+1 scenario, in vacuum, only two effective CP phases out of three contribute (see, \eg ~\cite{Gandhi:2015xza}). \cite{Berryman:2015nua} estimates 
the sensitivity of the DUNE experiment to the new angles and the two effective
phases in the 3+1 scenario. Adding to the 
analyses in \cite{Berryman:2015nua}, we show how the 
sensitivity of long-baseline experiments to CP violation 
and the mass hierarchy depends on the new mixing angles
and all three CP phases. Our work in this paper incorporates the fact that the presence of matter redefines the eigenstates, and quantitatively demonstrates that even those parameters
which do not affect the $\nu_{\mu}\rightarrow\nu_{e}$
oscillations in vacuum can have significant effects
in the presence of matter. In other words, it translates and extends, to the level of sensitivities, this important physics point  regarding the effect of matter made earlier in  \cite{Gandhi:2015xza}, which demonstrated it  at the level of probabilities and event rates.
In \cite{Agarwalla:2016mrc, Agarwalla:2016xxa},
the authors try to determine the potential of T2K, NO$\nu$A and 
DUNE to distinguish a CP-violating value of a given 
CP phase $\delta_{x}$ from 
a situation where the same phase $\delta_{x}$ has been assumed
to be CP conserving while leaving the other two phases free. 
In contrast to this, we determine the potential of the 
long-baseline experiments to distinguish between the 
situation where any one or more of the CP phases has 
a CP-violating value, and the situation where all the 
CP phases have CP-conserving values. In other words, 
to distinguish between the situation where CP is violated 
and the situation where it is not. We also try to provide an 
understanding of why the sensitivities to CP 
violation and the mass hierarchy behave as they do.

In Section \ref{Physics} below we define our mixing 
matrix parametrisation, describe our numerical simulation 
procedure and specify  the constraints that motivate the 
parameter ranges we use in the simulation. We also 
provide brief descriptions and   salient specifications 
for the long-baseline experiments considered in our 
work. Section \ref{results} gives the results on mass 
hierarchy and CP sensitivities in the presence of a 
sterile neutrino and discusses the reasons why they 
differ substantially from the 3+0 case. This section 
also  addresses the question of how large sterile-active 
mixing angles need to be before measurable effects 
show up at  long-baselines, and  how small they need 
to be such that their presence can be safely ignored 
when interpreting the results of experiments. 
Section \ref{summary} provides the conclusions and 
summarizes the results of the paper.

%====================================================================================
\section{Inputs and Calculational procedures for  the 3+1 scenario}
\label{Physics}
\subsection{The 3+1 mixing matrix}
We begin by specifying our  parameterisation for the PMNS matrix
in the presence of a sterile neutrino; 

\beq
U_{\text{PMNS}}^{3+1}=O(\theta_{34},
\delta_{34})O(\theta_{24},\delta_{24})O(\theta_{14})O(\theta_{23})O(\theta_{13},
\delta_{13})O(\theta_{12})
\eeq
Here,  in general, $O(\theta_{ij},\delta_{ij})$ is a rotation matrix
in the $ij$ sector with associated phase $\delta_{ij}$. For example,
\begin{equation*}
O(\theta_{24},\delta_{24}) = 
\begin{pmatrix}
1 & 0 & 0 & 0 \\
0 & \cos\tb & 0 & e^{-i\delta_{24}}\sin\tb \\
0 & 0 & 1 & 0 \\
0 & -e^{i\delta_{24}}\sin\tb & 0 & \cos\tb
\end{pmatrix};
O(\theta_{14}) = 
\begin{pmatrix}
\cos\ta & 0 & 0 & \sin\ta \\
0 & 1 & 0 & 0 \\
0 & 0 & 1 & 0 \\
-\sin\ta & 0 & 0 & \cos\ta
\end{pmatrix} \text{\etc}
\end{equation*}

In this parametrization, although the 
 the vacuum $\nu_\mu\rightarrow\nu_e$ oscillation probability is independent of the 3-4 mixing angle and the
associated CP phase,
the presence of matter 
brings about  a  dependence  on all mixing angles and phases. Specifically,
unlike the vacuum case, the 3-4 mixing angle and its associated phase are no
longer dormant, and the 3+1 electron neutrino appearance probability at long baselines exhibits a significant
dependence on them. Additionally, there  are interference terms that enter for non-zero values of the phases that are not necessarily small, especially in the presence of matter,  \eg, the term proportional to the sine of the sum of the phases $\delta_{13}$ and $\delta_{24}$ \cite{Gandhi:2015xza, 1742-6596-315-1-012015}.

\subsection{Simulation Procedure}
\label{SimulationResults}
\vspace{0.2cm}
In this section, we describe the details of the simulation technique
adopted in estimating the sensitivities and other  results obtained. 
We have used the  GLoBES \cite{Huber:2004ka, Huber:2007ji} software package
for performing all our analyses. For extending the simulation
to the 3+1 scenario, we used \cite{jkopp_snu, Kopp:2007ne}; which is an add-on to the default 
GLoBES software. Our assumptions regarding the values and ranges of the oscillation parameters for the 3+0 sector 
are as follows. 
\begin{itemize}
 \item  $\tx$ and $\ty$ are taken to be $33.48^\circ$ and $8.5^\circ$
respectively \cite{Gonzalez-Garcia:2014bfa}. 
\item $\delta m^2_{21}$ is taken to be $7.5\times10^{-5}~\rm{eV}^2$ 
while  $\delta m^2_{31}$ is set to be $2.457\times10^{-3}~\rm{eV}^2$
($-2.374\times10^{-3}~\rm{eV}^2$) for NH (IH) \cite{Gonzalez-Garcia:2014bfa}. 
\item The currently-allowed $3\sigma$ range on $\tz$ is 
$[38.3^\circ,53.3^\circ]$ with the best fit at $42.3^\circ (49.5^\circ)$ 
for NH (IH) \cite{Gonzalez-Garcia:2014bfa}. The $\tz$ best fit values 
from the global analyses \cite{Capozzi:2013csa, Forero:2014bxa} are 
somewhat different from \cite{Gonzalez-Garcia:2014bfa}. In this work, 
we make the simplifying assumption that 2-3 mixing is maximal; i.e. $\tz=45^\circ$. 
However, the conclusions we draw also apply to non-maximal
2-3 mixing.
\end{itemize}

It is anticipated that even if the 3+0 scenario is not realised in 
nature, the above values and ranges will still hold to a very 
good approximation.\footnote{\footnotesize Some of our early calculations showed
that the disappearance data {\it at the far detector} are less affected
by the active-sterile mixing angles compared to the appearance data.
Thus, the measurements that depend on $P_{\mu\mu}$, like
$\sin^22\tz$ or $|\delta m^2_{31}|$, are expected to change less with
the change of theoretical framework from 3+0 to 3+1. Likewise,
it was shown in \cite{Esmaili:2013yea} that $\ty$ measurements at 
the reactor neutrino experiments will be robust even if there
are sterile neutrinos. Also see \cite{Berryman:2015nua} regarding this.}

We draw information regarding the value of $\delta m^2_{41}$ from
\cite{Kopp:2013vaa}, which does a combined analysis of the global data. 
The best fit of $\delta m^2_{41}$ in the 3+1 scheme is found to be 
$0.93~\rm{eV}^2$ (Table 8 of \cite{Kopp:2013vaa}).
In our work,  we assume $\delta m^2_{41}$ to be $+1~\rm{eV}^2$\footnote{\footnotesize The choice $\delta m^{2}_{41} \sim1\textrm{eV}^{2}$ is made because it is a convenient benchmark. We have checked that our results and conclusions remain qualitatively valid for the mass splitting range $0.1-10$ eV$^2$. Numerical differences in an important calculated quantity like the event rate, for instance,  between our benchmark choice and the upper and lower bounds of this range are of the order of $10\%$, and will bring about corresponding changes in calculations of sensitivities and other quantities calculated in our paper.}
along with 
\beq
\delta m^2_{41} \sim \delta m^2_{42} \sim \delta m^2_{43} >> |\delta m^2_{31}|\sim|\delta m^2_{32}|
>> \delta m^2_{21}
\eeq

Our assumed ranges for the sterile sector mixing angles corresponding 
to the 3+1 scenario draw upon current constraints and are as follows.
Note that we derive the constraints correlated with the new mass-squared
difference; i.e., the values of mixing angles that are compatible with $\delta m^2_{41} \sim~1~\rm{eV}^2$ are chosen.
\begin{itemize}\itemsep 0.05cm
\item Measurements at the Daya Bay experiments put constraints on the 
effective mixing angle in the electron anti-neutrino disappearance channel.
This effective mixing angle is the same as $\ta$ under the choice of 
PMNS parameterisation in this work. Based on \cite{An:2014bik},
we assume $\ta\leq13^\circ$ at 95\% C.L. Slightly tighter constraints
are available from the BUGEY experiment; 
but at 90\% C.L. \cite{Declais:1994su}
\item The strongest constraints on the 2-4 mixing angle can be
derived from the IceCube data \cite{jones_icecube2016}. 
With their current data, only $\tb\leq7^\circ$ can be allowed at 99\% C.L.
\item The MINOS experiment with its observed charged-current and neutral-current events spectra can  constrain the 3-4 mixing angle. From \cite{Adamson:2011ku}, we have $\tc\leq26^\circ$ at 90\% C.L. 
\end{itemize}

 We also  vary 
 $\da$, $\db$ and $\dc$ for 3+1 and $\dcp$ for 3+0 over the full possible 
 range of $[-180^\circ, 180^\circ]$. Finally, the fluxes we use are identical to those used in \cite{Bass:2013vcg}\footnote{\footnotesize These differ slightly in intensity and peak value from the present version used by the DUNE collaboration, but these differences do not affect our conclusions in any substantive manner.}. Details regarding the calculation of Poissonian $\chi^2$,
treatment of systematic uncertainties etc. can be found in 
\cite{Huber:2004ka, Huber:2007ji}.

%====================================================================================
\subsection{Details of Experiments}
\label{Experiments}

{\bf{T2K and T2HK}}

The Tokai to Kamioka (T2K) experiment is an  ongoing
neutrino experiment in Japan whose main goals are to observe
$\nu_{\mu}\rightarrow\nu_{e}$ oscillations and to measure $\ty$.
It may, however, be in a position to provide hints on CPV and the hierarchy, especially when its data are used in conjunction with those of other experiments.
Neutrino beams generated at the J-PARC accelerator facility 
in Tokai are directed towards a 22.5 kton 
water \v{C}erenkov detector placed in Kamioka, 295 km away 
at a $2.5^\circ$ off-axis angle \cite{Itow:2001ee}. The $\nu_{\mu}$ 
beam peaks sharply at 0.6 GeV, which is very close to the 
first oscillation maximum of the $\nu_\mu\rightarrow\nu_e$ appearance probability, $\pme$. The flux falls off quite rapidly, 
such that it is negligible at energies greater than 1 GeV. 
The   beam 
power is 750 kW, with a proton energy of 30 GeV, for runs in both the 
$\nu$ and $\bar{\nu}$ modes. 
 Combining both runs,
the experiment will gather a total exposure of $\sim 8\times10^{21}$ 
protons on target (POT).
 The neutrino flux is monitored by the 
near detectors, located 280 m away from the point of neutrino 
production.  Details regarding the detector efficiencies and 
background events used in our work  have been taken from \cite{Huber:2004ka}.

The T2HK experiment \cite{Abe:2015zbg}, in essence, is a 
scaled-up version of the T2K experiment. It will accumulate 
$1.56\times10^{22}$ protons on target with a 30 GeV proton beam.
The detector size is expected to be 25 times the T2K detector. For both T2K and T2HK, we have assumed 2.5\% (5\%) signal and  20\% (5\%) background normalisation errors in $\nu_{\mu}$ ($\nu_{e}$) signal. All details regarding signal and background
events and detector efficiencies for T2HK are taken from \cite{Abe:2015zbg, Ishitsuka:2005qi}.
 
{\bf{NO$\nu$A}}

The NuMI{\footnote{Neutrinos at the Main Injector.} }
Off-axis $\nu_{e}$ Appearance experiment (NO$\nu$A) \cite{Ayres:2007tu} 
is an ongoing long-baseline super-beam experiment 
in the US. It aims to determine the 
mass hierarchy, $\ty$ , the octant of $\tz$, and perhaps leptonic CP-violation, by the measurement of 
$\nu_{\mu}\rightarrow\nu_{e}$ oscillations. The source of $\nu_{\mu}$ 
is  the Fermilab's NuMI beamline. A 14 kton Totally Active 
Scintillator Detector (TASD) is placed in Ash River, Minnesota, 
which is 810 km away at an off-axis angle of 14 mrad ($0.8^\circ$).
This off-axis narrow-width beam peaks at 2 GeV.
The experiment is scheduled to run for 3 years in 
$\nu$ mode and 3 years in $\anu$ mode
with a NuMI beam power of $0.7$ MW and 120 GeV
proton energy, corresponding to $6\times 10^{20}$ 
POT per year. A 0.3 kton near detector is located 
at the Fermilab site. We assume 5\% (2.5\%) signal normalisation error for $\nu_{e}$ ($\nu_{\mu}$) signal. Assumed background error is 10\%. The details of signal and background
events and the detector efficiencies have been taken from
\cite{Agarwalla:2012bv}.

{\bf{DUNE}}

 DUNE, a future experiment scheduled to come online $\sim2025$, (with specifications very similar to
LBNE \cite{Hewett:2012ns, Adams:2013qkq}), will be located 
in the United States. It is a super-beam 
experiment with the main aim of establishing or refuting the existence 
of CPV in the leptonic sector. In addition to this primary goal, it 
will also be able to resolve  the mass hierarchy 
and shed light on the octant of $\tz$. The ${\nu_{\mu}(\bar{\nu}_{\mu}})$ super-beam 
will originate at  Fermilab. The primary beam simulation assumes 
a 1.2 MW - 120 GeV proton beam that will deliver $10^{21}$ 
protons-on-target (POT) per year. A 35-40 kt Liquid Argon (LAr) 
far-detector will be housed in the Homestake mine in South Dakota, 1300 km
away\footnote{In addition to this, there is a proposal to install a near detector\cite{Acciarri:2016ooe, Choudhary-nd}, which among other physics goals can also constrain the parameter space for the 3+1 scenario, as recently discussed in \cite{Choubey:2016fpi}.}. The experiment plans to have a total of 10 years of running, divided
equally between neutrinos and anti-neutrinos,  corresponding  to a total
exposure of $35\times10^{22}$ kt-POT-yr.  Other experimental details,
such as signal and background definitions as well as the detector efficiencies 
taken in this work are the same as those in \cite{Bass:2013vcg}, except with
the  difference that we have not considered tau events in the 
backgrounds. We assume a 5\% signal normalisation error and a 10\% 
background normalisation error. The detector efficiencies for both
$\pme$ and $\pmebar$ events are close to 80\% with somewhat less
efficiency for $\pmebar$.

%====================================================================================
\section{Results and Discussion}
\label{results}

\subsection{Sensitivity to CP violation in the presence of a single sterile neutrino}
\label{CPviolation}

\begin{figure}[H]
\centering
\includegraphics[width=0.49\textwidth]{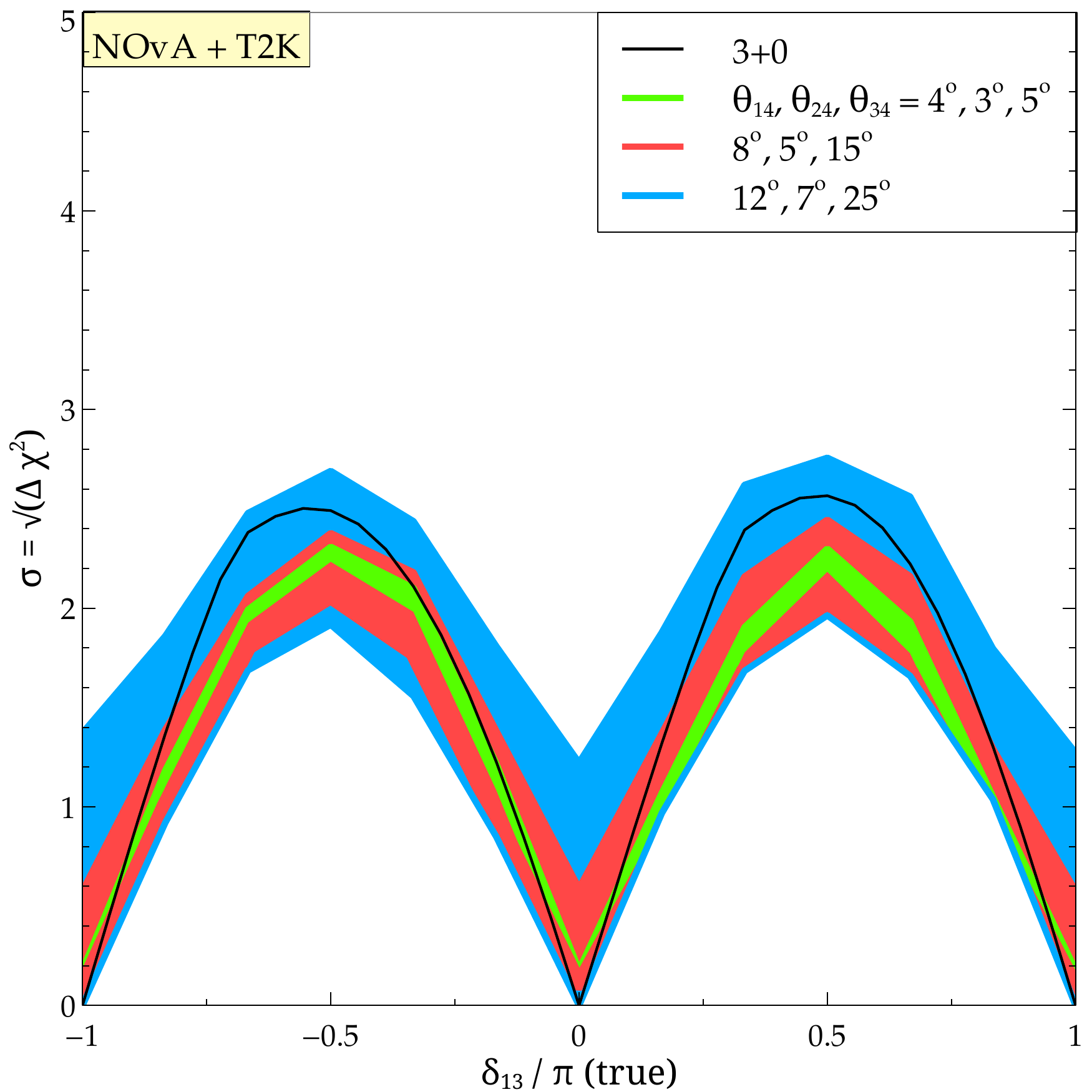}
\includegraphics[width=0.49\textwidth]{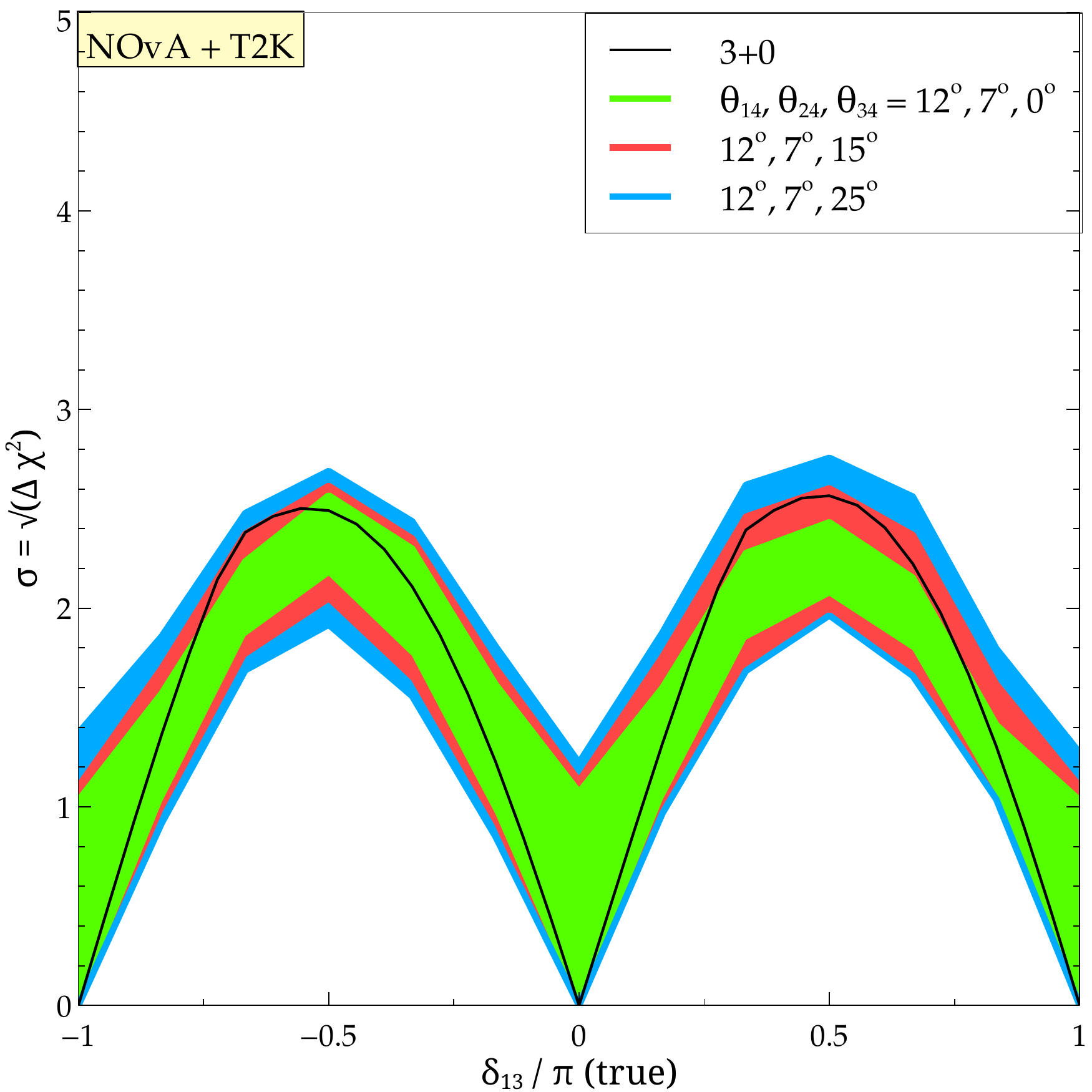}
\caption{\footnotesize{Sensitivity to CP violation as a function of the true
CP violating phase $\da$ for the combined data from T2K and NO$\nu$A. Different colors 
correspond to different choice of true $\ta, \tb, \tc$ as shown in the key. Variation
of true $\db$ and $\dc$ results in the colored bands which show the minimum
and maximum sensitivity that can be obtained for a particular $\da$. The black
curve corresponds to sensitivity to CP violation in 3+0. Left panel: Shows the
effect as all the three active-sterile mixings are increased. Right panel: Shows
the effect of the 3-4 mixing when the true $\ta$ and $\tb$ have been fixed at 
$12^\circ$ and $7^\circ$ respectively for all three bands.}}
\label{CPnovat2k}
\end{figure}

\begin{figure}[H]
\centering
\includegraphics[width=0.49\textwidth]{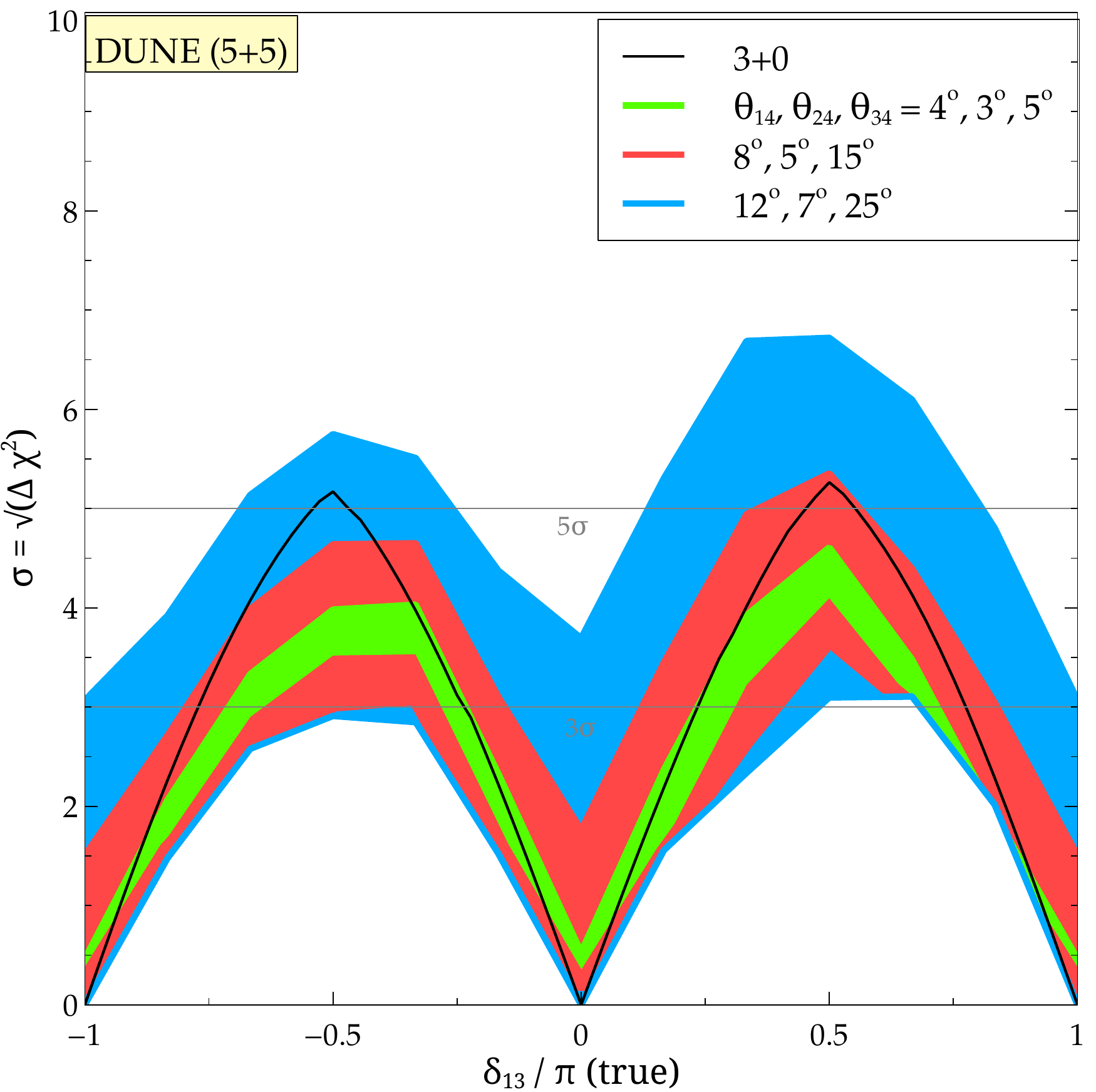}
\includegraphics[width=0.49\textwidth]{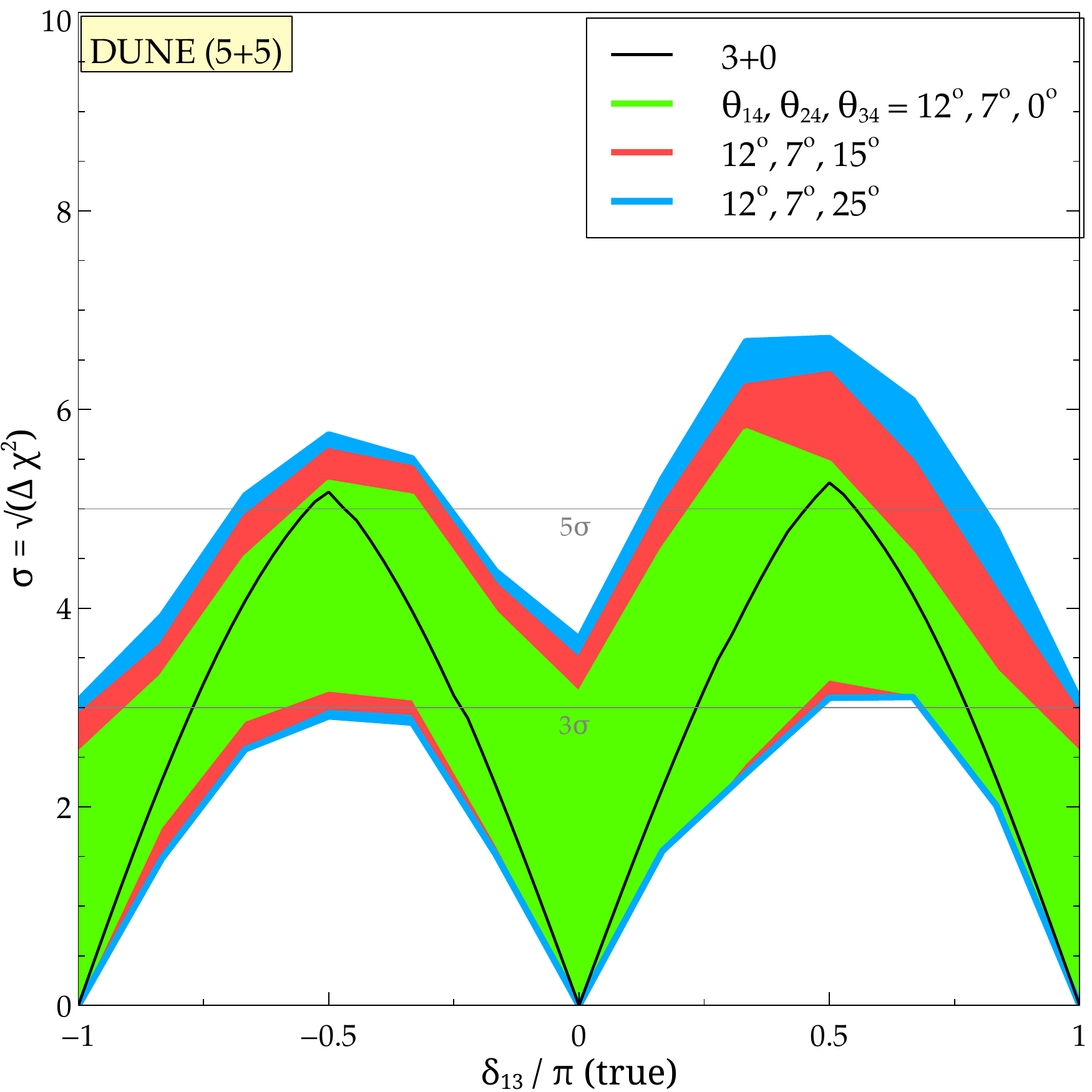}
\caption{\footnotesize{
Similar to Fig.\ \ref{CPnovat2k} but for DUNE. The hierarchy is assumed to be normal.
}}
\label{CPdune}
\end{figure}

\begin{figure}[H]
\centering
\includegraphics[width=0.49\textwidth]{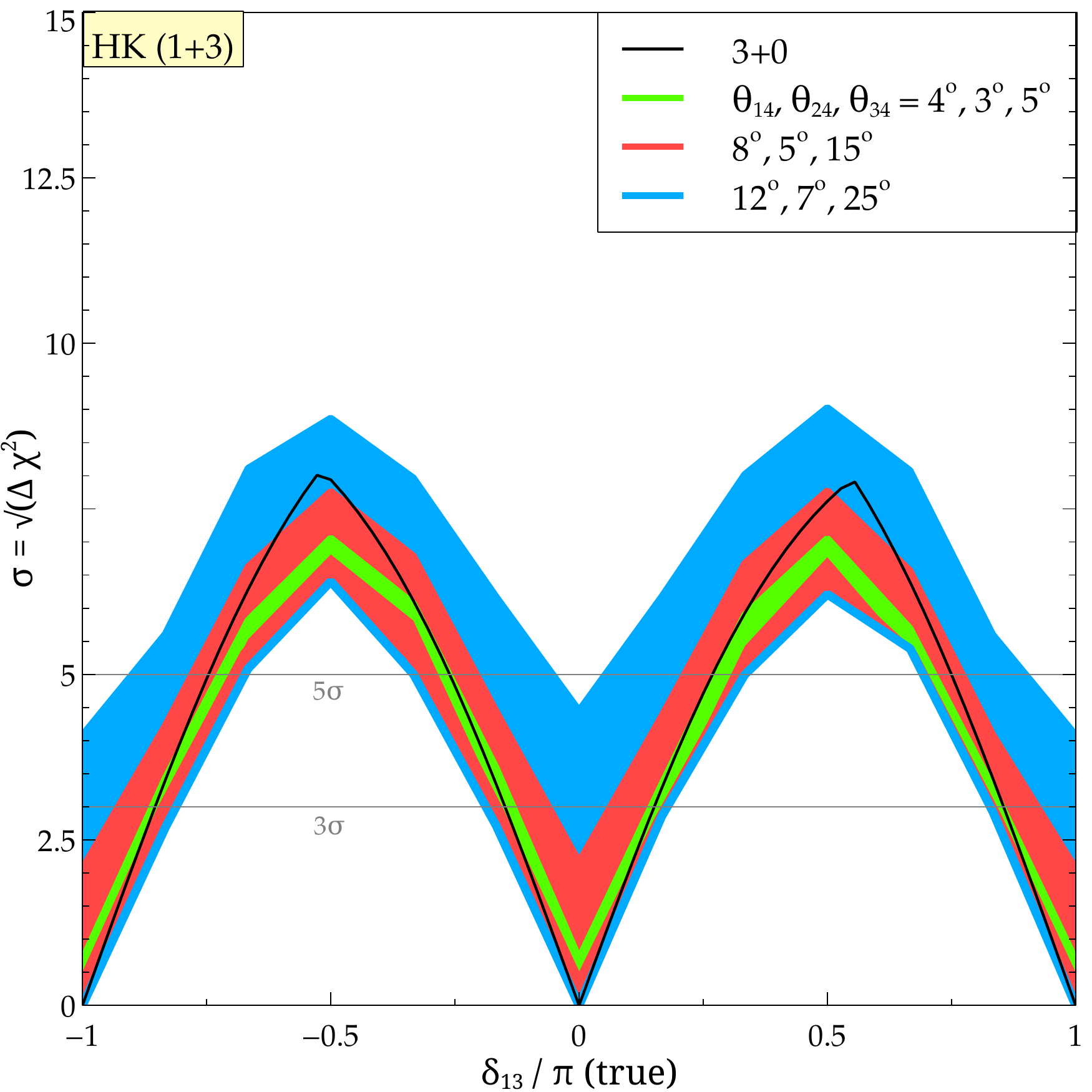}
\includegraphics[width=0.49\textwidth]{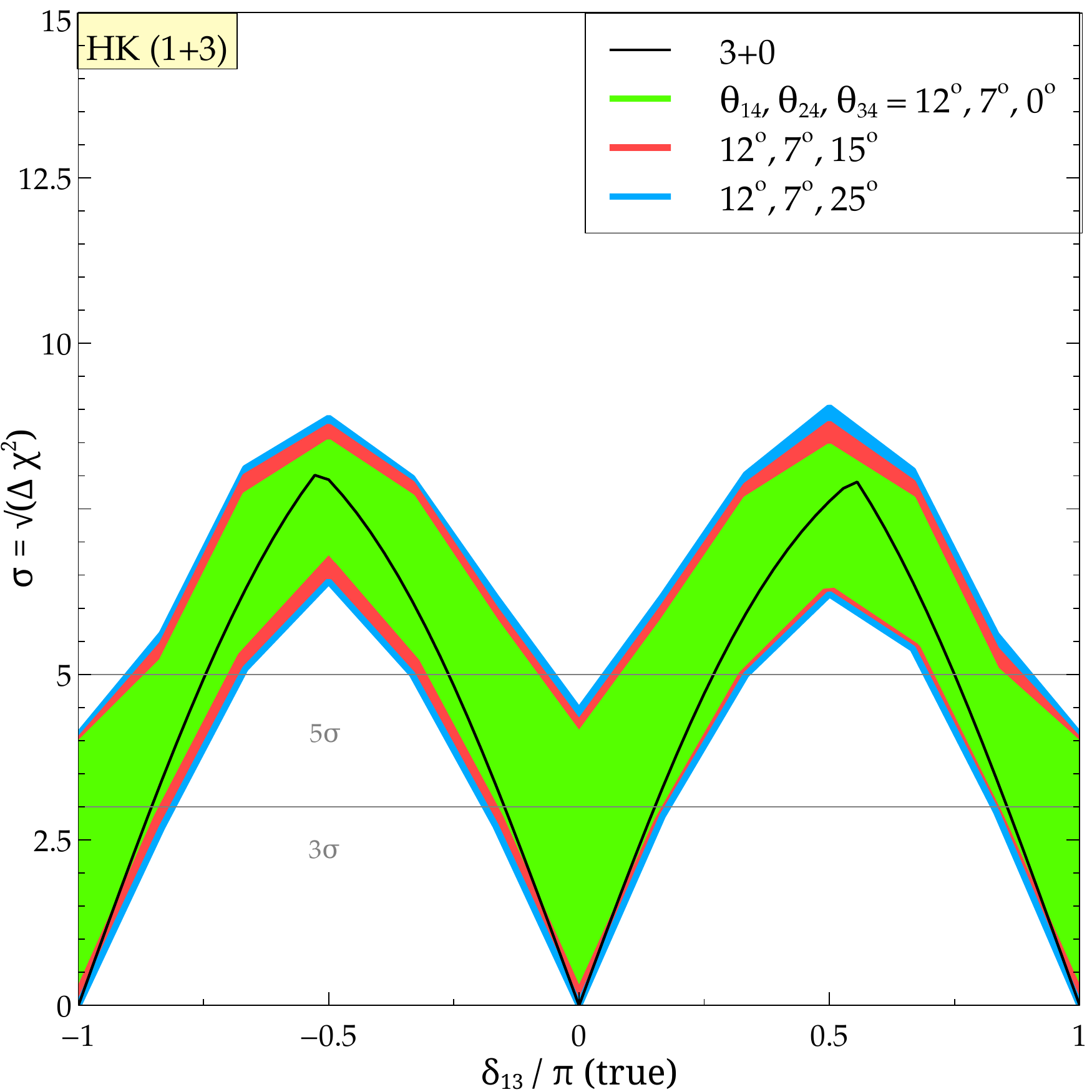}
\caption{\footnotesize{
Similar to Fig.\ \ref{CPnovat2k} but for T2HK.
}}
\label{CPhk}
\end{figure}

\begin{figure}[H]
\centering
\includegraphics[width=0.49\textwidth]{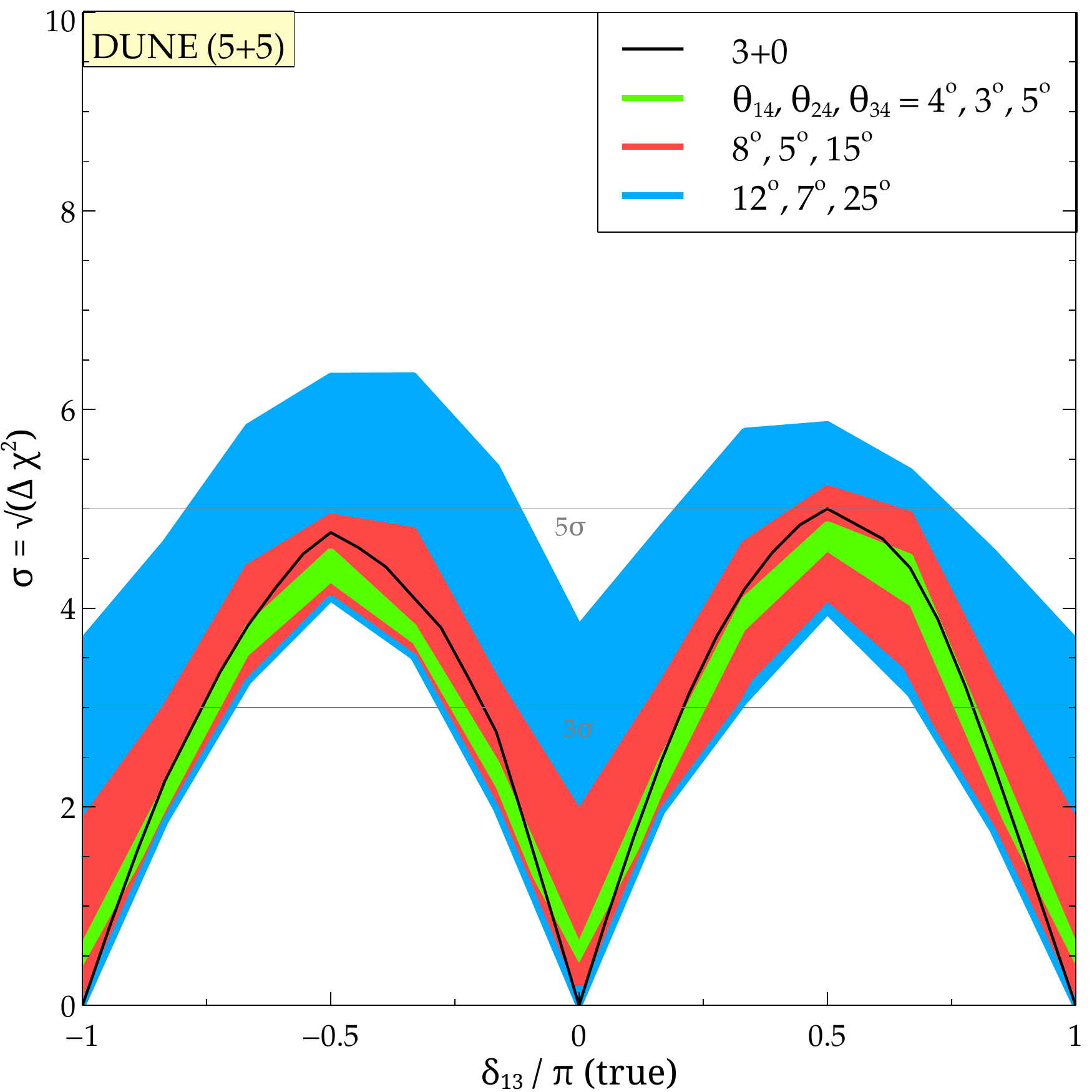}
\includegraphics[width=0.49\textwidth]{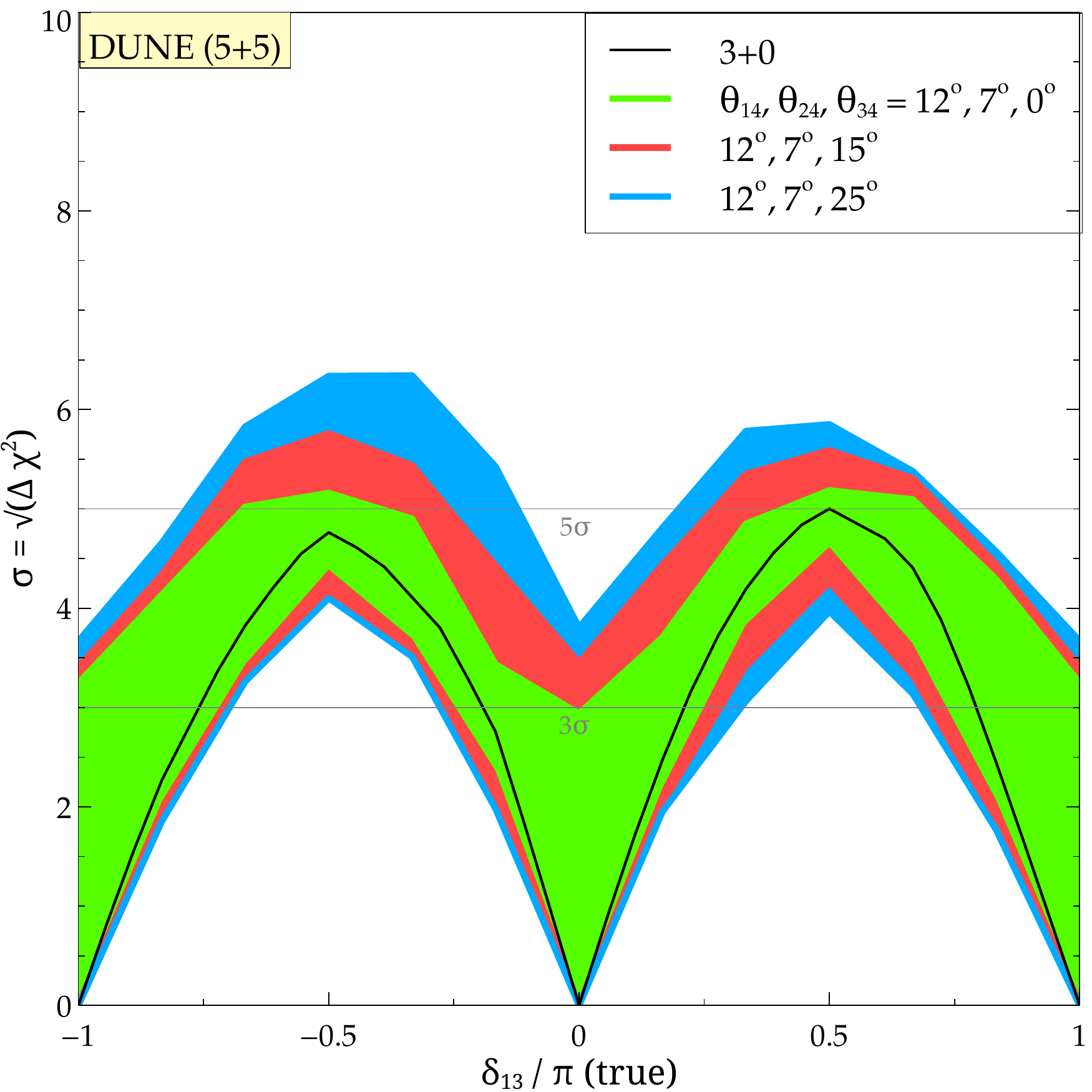}
\caption{\footnotesize{
Similar to Fig.\ \ref{CPdune} but for inverted hierarchy (IH).
}}
\label{CPdune_ih}
\end{figure}

In this section, we show the sensitivities of various long-baseline
experiments to leptonic CP violation in the 3+1 scenario. Results for the 3+0 case are provided for comparison.
 Figs. 
\ref{CPnovat2k} to \ref{CPdune_ih} show the sensitivity to excluding 
the CP conserving values as a function of the 3+1 oscillation parameters.

We show results for three sets of the active-sterile mixing angles 
$\ta, \tb, \tc$. These are chosen to be  $(4^\circ, 3^\circ, 5^\circ)$,  $(8^\circ, 5^\circ, 15^\circ)$ 
and $(12^\circ, 7^\circ, 25^\circ)$. The {\it data} are simulated assuming 
the above three sets of mixing angles and various choices of the 
three CP phases lying in $[-180^\circ, 180^\circ]$. In the {\it fit}, we 
consider the 8 possible CP conserving combinations of $\da, \db, \dc$ 
where each phase could either be 0 or $180^\circ$. We minimise the 
$\dxx$ over these 8 test CP conserving cases; and 
over a relatively fine grid of test $\ta, \tb, \tc$ samples in the allowed range in the fit,
so as to account for the lack of information regarding active-sterile mixings.
This gives us a $\dxxmin$ as a function of the true parameters.
% Both for simulating data and in the fit, we assume the hierarchy
% to be normal only.
In creating Figs.\ \ref{CPnovat2k} to \ref{CPhk}, both for simulating data and in the fit, we assume the hierarchy to be normal only (we will consider an inverted hierarchy shortly). 
We did not marginalise over the 3+0 parameters except $\delta_{13}$\footnote{Our 
results show that a close to $5\sigma$ determination 
of hierarchy is very likely with the DUNE experiment, even in the 
3+1 paradigm. Among other 3+0 oscillation parameters, marginalisation
over $\tz$ may be important when the non-maximal true values like
the ones in lower octant or higher octant are considered because of
octant-related degeneracies. Since,  disappearance 
($P_{\mu\mu}$) data fixes $\sin^22\tz$ very accurately, 
marginalisation over test $\tz$ is not necessary for true $\tz=45^\circ$.}. 
For a particular true $\da$, we show the maximum and the minimum 
$\dxxmin$ that can be obtained corresponding to a variation of the other 
two true CP phases $\db$ and $\dc$. For 3+0, the situation is simpler, 
where we contrast a true $\dcp$ against $\dcp=0$ and $\dcp=\pi$ in the fit.

It can be seen from Figs.\ \ref{CPnovat2k} to \ref{CPhk} that the existence 
of sterile neutrinos can significantly affect the CPV discovery  potential of  long-baseline
experiments. This violation can originate in any of 
the three phases and not just $\da$. When the 
active-sterile mixings are small, the general trend visible in the figures is that the sensitivity
to CP violation of the experiment will be decreased compared to what we
would expect in the 3+0 scenario. However, for sufficiently large mixings, the sensitivity
spans  both sides of the 3+0 curve; and hence, depending on the true 
value of the other phases - $\db$ and $\dc$, the sensitivity to CP violation
can be greatly amplified. We  observe that the chosen value of true 
$\tc$ significantly affects the sensitivity to CP violation, especially for  
DUNE, where the matter effects are large. We also note that there can be  significant amplification of CPV sensitivity for regions of $\da$ where one expects little or none in the 3+0 scenario. 

Similar observations can also be made when the true hierarchy is considered to be  inverted. As an example, we have shown in Fig.\ \ref{CPdune_ih} the CPV sensitivity results for DUNE (similar to Fig.\ \ref{CPdune}) for inverted hierarchy and the features are qualitatively similar. 

Overall, if sterile mixing angles are not tiny, the 3+1 (or 3+n) scenario, if realized in nature,  makes the observation of generic CPV {\it{per se}} significantly more likely than the 3+0 case, although it makes the determination of the phase (or phases) in which such violation originates much harder (we address this point in greater
detail in Subsection \ref{CPsource}).

In order to understand the behavior of sensitivities to CP violation and later the 
mass hierarchy, we note  the role of the following competing effects in the calculation of $\dxxmin$\footnote{Similar effects can also guide the value of $\dxxmin$ in the context of other new physics effects such as NSI and this has been illustrated in \cite{Masud:2016bvp, Masud:2016gcl}.}:
\begin{enumerate}
\item The parameter space for the 3+1 scenario consists of $\ta, \tb, \tc, 
\db, \dc$, and $\delta m^{2}_{41}$, in addition to the standard 3+0 parameters. 
While generating the $\dxxmin$ curve for 3+0 scenario (black), we marginalized only over the test parameter $\da$ (\ie,~\ 
$\dcp$); however, for the 3+1 curves, marginalization was carried over the five additional test parameters $\ta, \tb, 
\tc, \db, \dc$ in addition to $\da$. Hence, the parameter space for 3+1 case is a substantially larger
superset of the 3+0 parameter space. In general, from a statistical point of view, marginalization over a larger space of test parameters
tends to bring down the value of $\dxxmin$. 
\item As the true values of the 3+1 mixing angles ${\ta, \tb, \tc}$ increase, the impact due to the variation in the true values of the associated phases ($\db, \dc$) also increases. This results in substantially broadening the $\dxxmin$ bands (for 3+1) in both directions as the true values of ${\ta,\tb,\tc}$ increase from ${4^{\circ}, 3^{\circ}, 5^{\circ}}$ to ${12^{\circ}, 7^{\circ}, 25^{\circ}}$,- thereby making the $\dxxmin$ (for 3+1 scenario) sometimes even larger than the 3+0 $\dxxmin$.
\end{enumerate}
When the active-sterile mixing angles (true $\ta, \tb, \tc$) are small, 
effect (2) is small. However, the statistical effect (effect (1) above) stemming from  marginalization over five added parameters is significant and reduces the sensitivity in general.  Consequently, $\dxxmin$ tends to decrease for small values of the 
true sterile mixing angles. On the other hand, when the mixing angles (true $\ta, \tb, \tc$) 
increase, they concurrently amplify the effect of the CP violating phases. Consequently, effect (2) plays a correspondingly   important role, and tends to increase $\dxxmin$ overall. These features have been explained in 
greater detail in Subsection \ref{Totaleventsrates} using the DUNE total 
neutrino and anti-neutrino event rates as an example.

 \subsection{Mass hierarchy}
 \label{MassHierarchy}

In this section, we re-evaluate the sensitivities of the long-baseline 
experiments to the neutrino mass hierarchy, 
assuming the 3+1 scenario.  We first show  results assuming 
the normal hierarchy. The simulation procedure followed 
here is the same as that described  in Sec. \ref{CPviolation} except for 
the following important  differences:

\begin{figure}[H]
\centering
\includegraphics[width=0.49\textwidth]{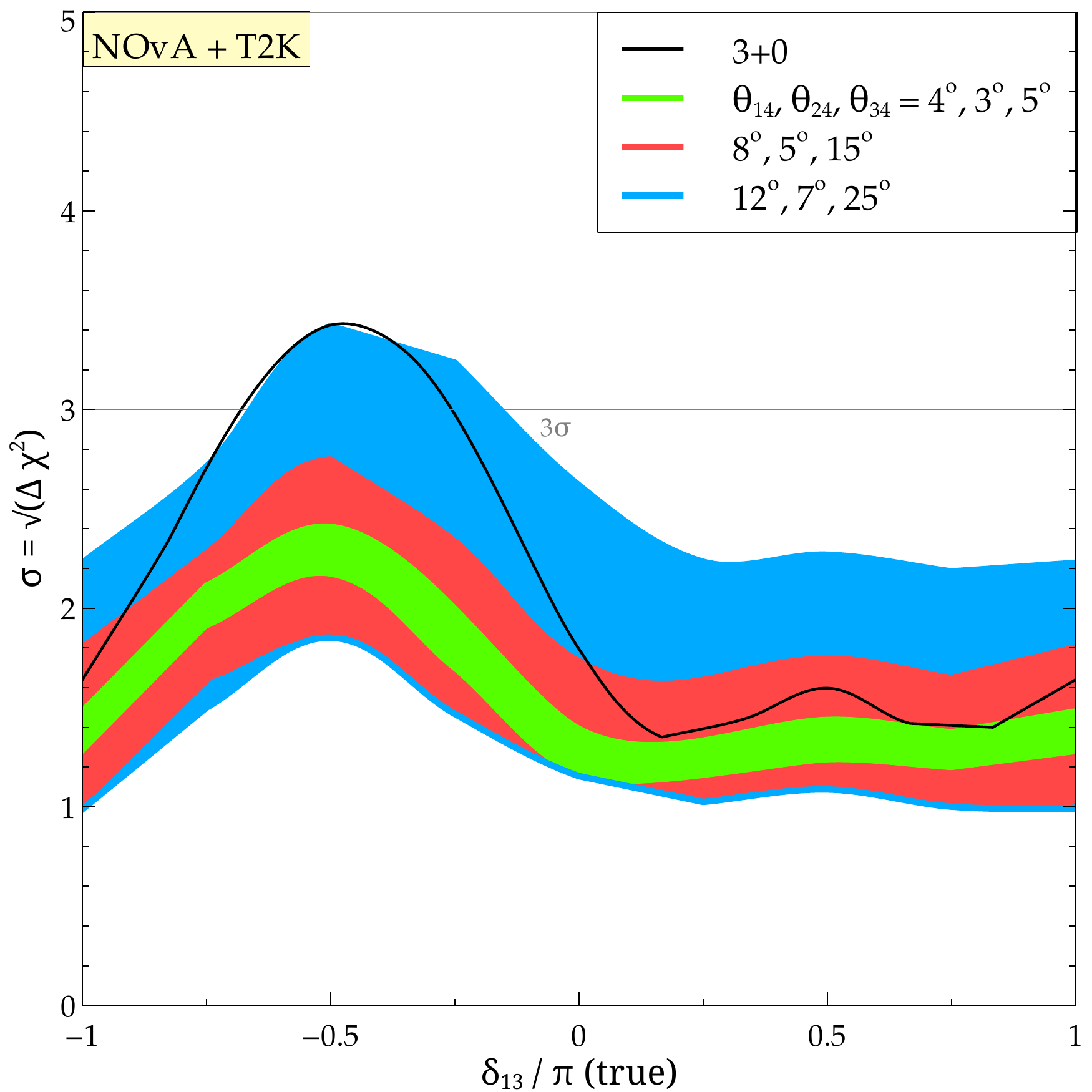}
\includegraphics[width=0.49\textwidth]{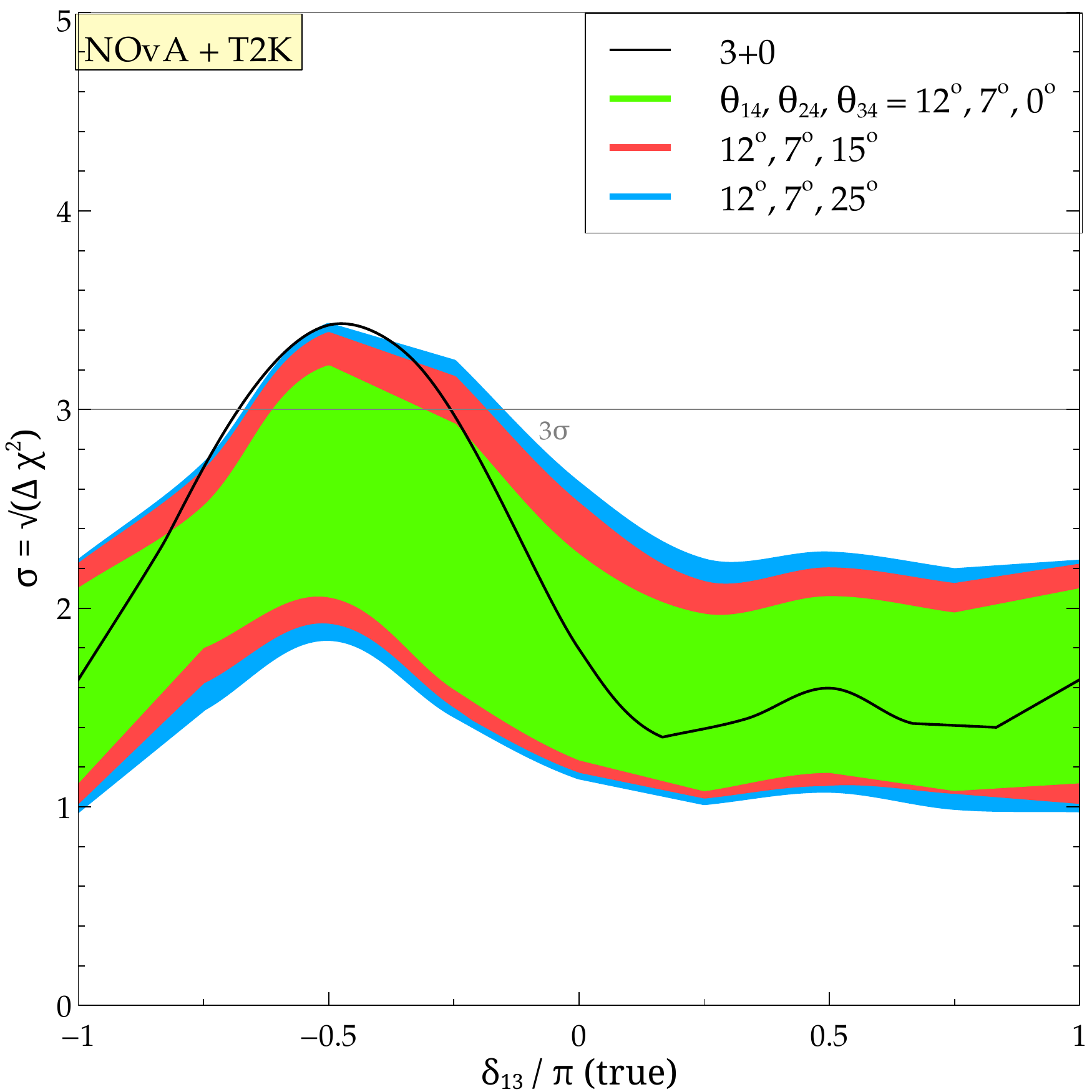}
\caption{\footnotesize{Sensitivity to Mass hierarchy as a function of the true
CP violating phase $\da$ for the combined data from T2K and NO$\nu$A. Different colors 
correspond to different choice of true $\ta, \tb, \tc$ as shown in the key. Variation
of true $\db$ and $\dc$ results in the colored bands which show the minimum
and maximum sensitivity that can be obtained for a particular $\da$. The black
curve corresponds to sensitivity to the hierarchy in 3+0. Left panel: Shows the
effect as all the three active-sterile mixings are increased. Right panel: Shows
the effect of the 3-4 mixing when the true $\ta$ and $\tb$ have been fixed at 
$12^\circ$ and $7^\circ$ respectively for all three bands.}}
\label{MHnovat2k}
\end{figure}

\begin{figure}[t]
\centering
\includegraphics[width=0.49\textwidth]{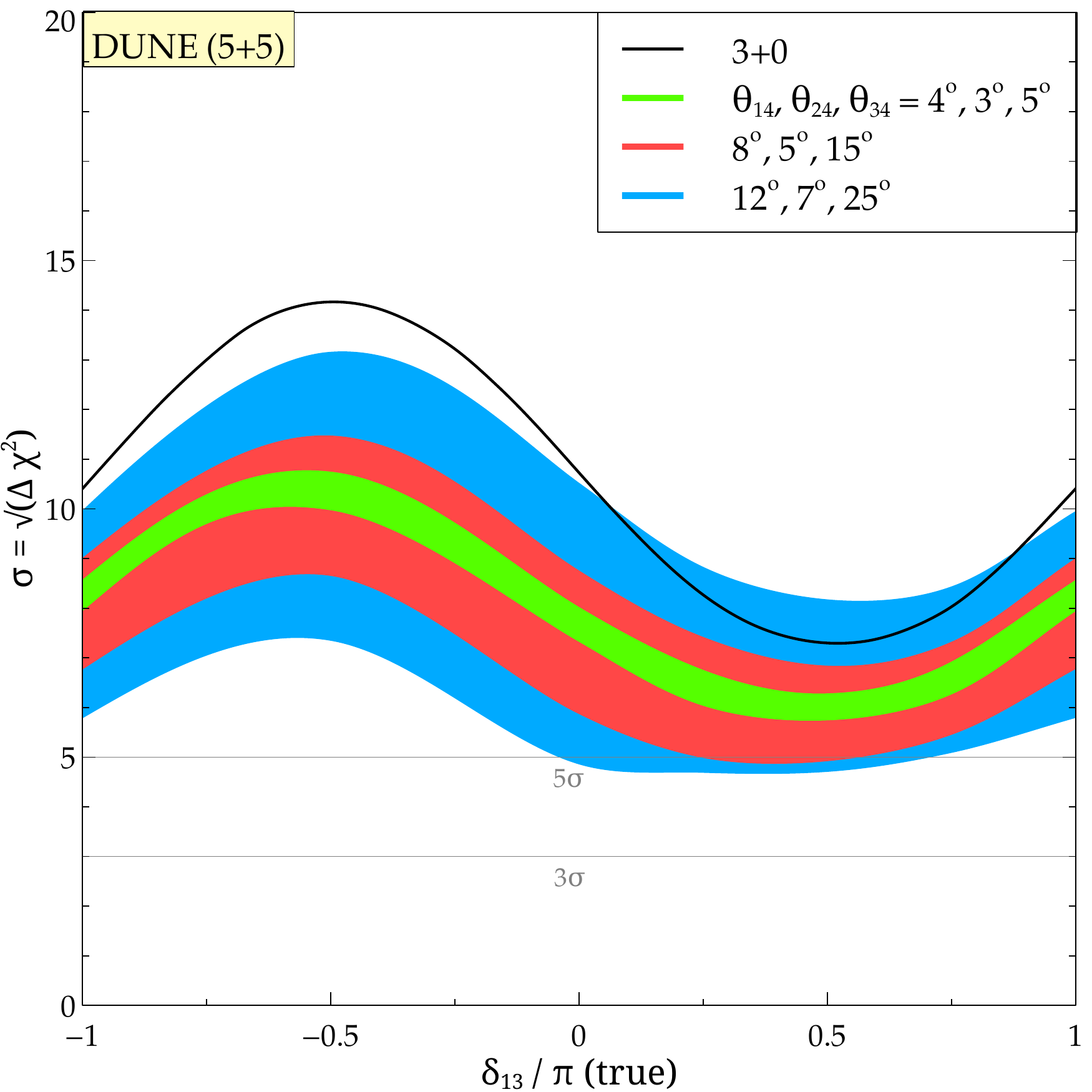}
\includegraphics[width=0.49\textwidth]{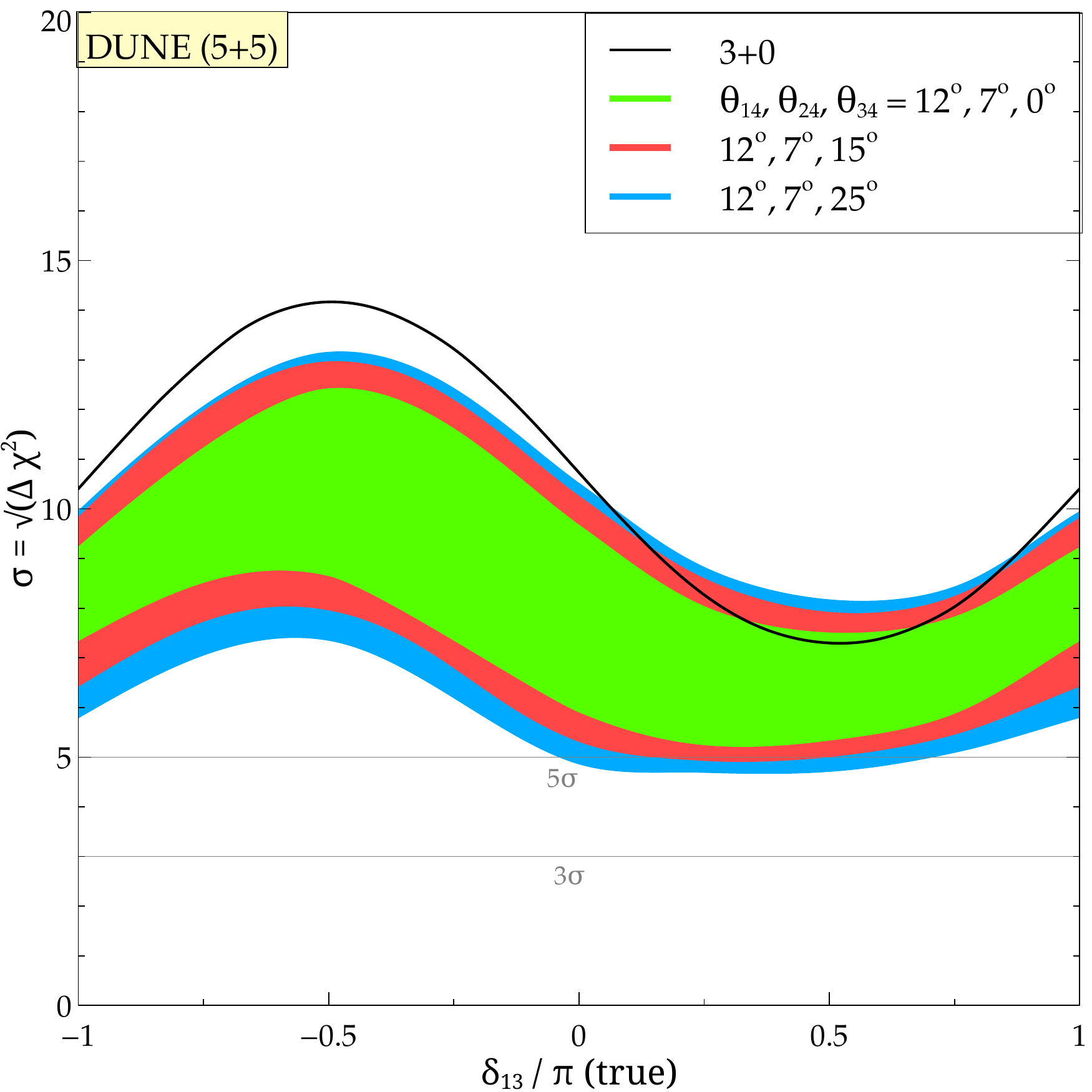}
\caption{\footnotesize{
Similar to Fig.\ \ref{MHnovat2k} but for DUNE. Normal hierarchy was considered to be true. 
}}
\label{MHdune}
\end{figure}

\begin{itemize}
\item In the fit, we assume the hierarchy to be inverted.
\item For  the CP violation sensitivities, in the 
fit, we had considered only those combinations of test values of $\da,\db,\dc$ which were CP conserving.
 For hierarchy determination,  however, we have varied these test CP 
phases in their full allowed range $[-180^\circ,180^\circ]$ and 
have marginalised over them.
\end{itemize}
Note that, as in the previous section, we did not vary the test 3+0 parameters other than the CP phase 
in the fit. We show the results in Figs. \ref{MHnovat2k} and
\ref{MHdune} as a function of the true 3+1 parameters.  For the 
combined results from T2K and NO$\nu$A, it can be seen 
in Fig. \ref{MHnovat2k} (left panel) that there emerges the possibility of 
significant improvement in the hierarchy sensitivity
compared to 3+0 (shown by black line) in the unfavourable 
regions of true $\da$. The extent of this enhancement is,  of course, dependent on the 
true values of the active-sterile oscillation parameters. Fig. 
\ref{MHnovat2k} (right panel) shows the dependence of sensitivity
on the  $\tc$ mixing angle - the effects of which are noticeable, although not large,  even 
for baselines where matter effects are not very substantial. 

\begin{figure}[t]
\centering
\includegraphics[width=0.49\textwidth]{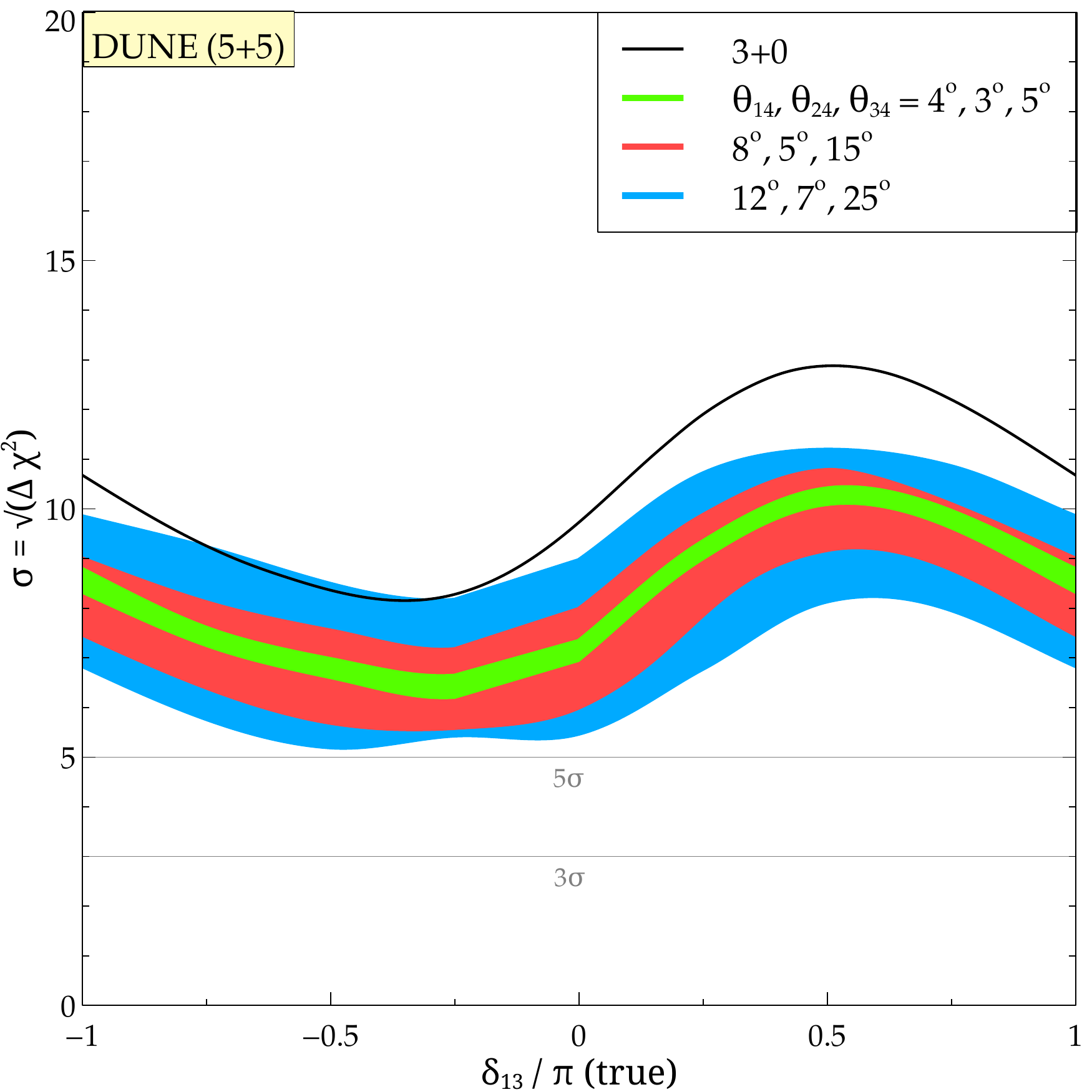}
\includegraphics[width=0.49\textwidth]{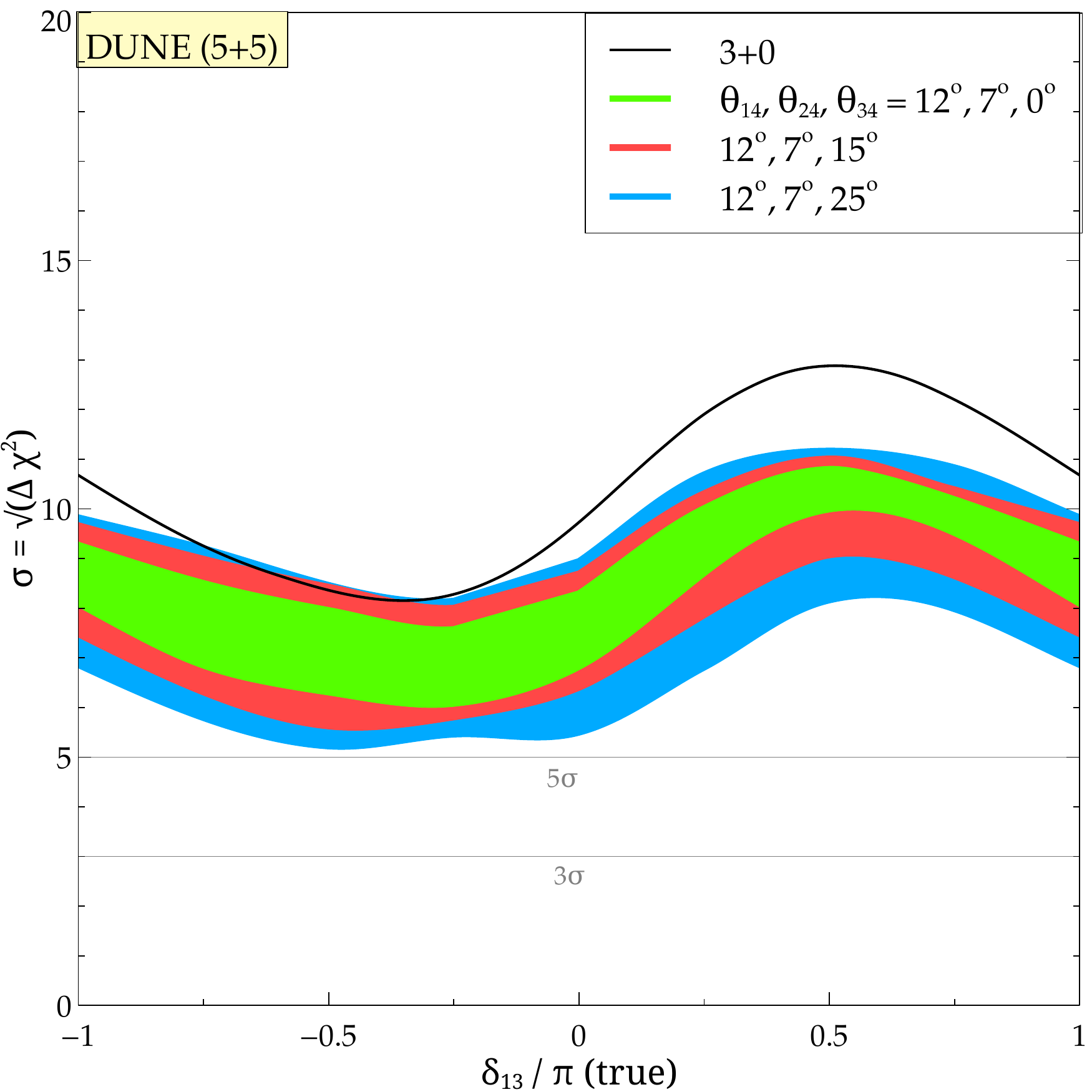}
\caption{\footnotesize{
Similar to Fig.\ \ref{MHdune} but when the true hierarchy is considered to be  inverted.
}}
\label{MHdune_ih}
\end{figure}

In the case of 
DUNE, we find that the 3+1 sensitivities are usually below 3+0
sensitivities\footnote{\footnotesize As mentioned in the beginning of the present subsection \ref{MassHierarchy}, the test CP phases have been marginalized over their full allowed range $[-180^{\circ}, 180^{\circ}]$ while calculating the $\dxxmin$ for mass hierarchy; whereas for CP violation sensitivity the marginalization was carried over the CP-conserving values - $0^{\circ}$ and $180^{\circ}$ only. This enhanced range of marginalization for mass hierarchy leads to a larger statistical effect (see effect (1) in the previous subsection), leading to a generally reduced $\dxxmin$ for 3+1, compared to the 3+0 case.} except for a small region of parameter space around
true $\da=90^\circ$, as can be observed in both the panels of Fig. \ref{MHdune}.
The $\tc$-dependence of sensitivities is somewhat more
pronounced for DUNE,  as is evident from 
the right panel of Fig. \ref{MHdune}. It should be noted that except for
a small fraction of parameter space around true $\da=90^\circ$,
the sensitivity stays above $5\sigma$ C.L.

In Fig.\ \ref{MHdune_ih}, we have shown the mass hierarchy sensitivity of DUNE when the true hierarchy is inverted. The positions of the peaks and troughs of the sensitivities are opposite to those in Fig.\ \ref{MHdune}. The general features of the sensitivities are qualitatively similar to those corresponding to a normal true hierarchy (as in Fig.\ \ref{MHdune}), except for the following small difference:- the hierarchy sensitivity bands in Fig.\ \ref{MHdune_ih} in the presence of a sterile neutrino are slightly narrower than those in Fig.\ \ref{MHdune}. The sensitivities in the presence of a sterile neutrino remain above the $5\sigma$ C.L.

\subsection{Using total event rates to understand the DUNE sensitivity to Hierarchy and CPV}
\label{Totaleventsrates}

\begin{figure}[H]
\centering
\includegraphics[width=0.35\textwidth]{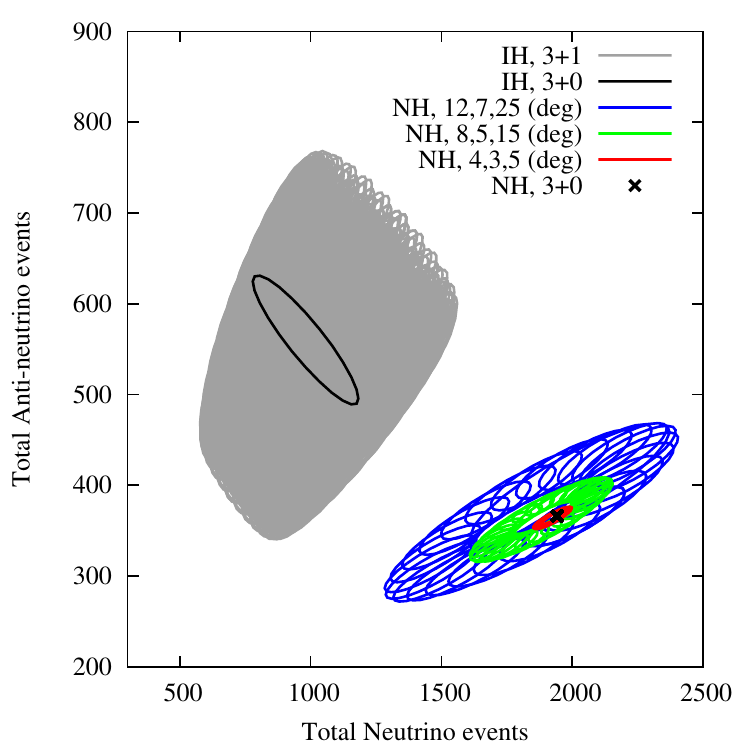}
\includegraphics[width=0.35\textwidth]{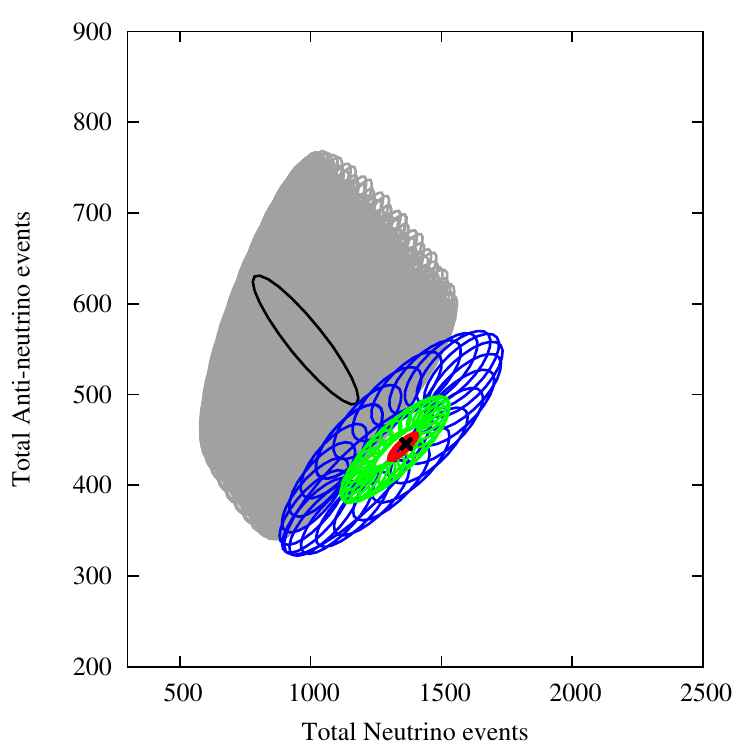}
\includegraphics[width=0.35\textwidth]{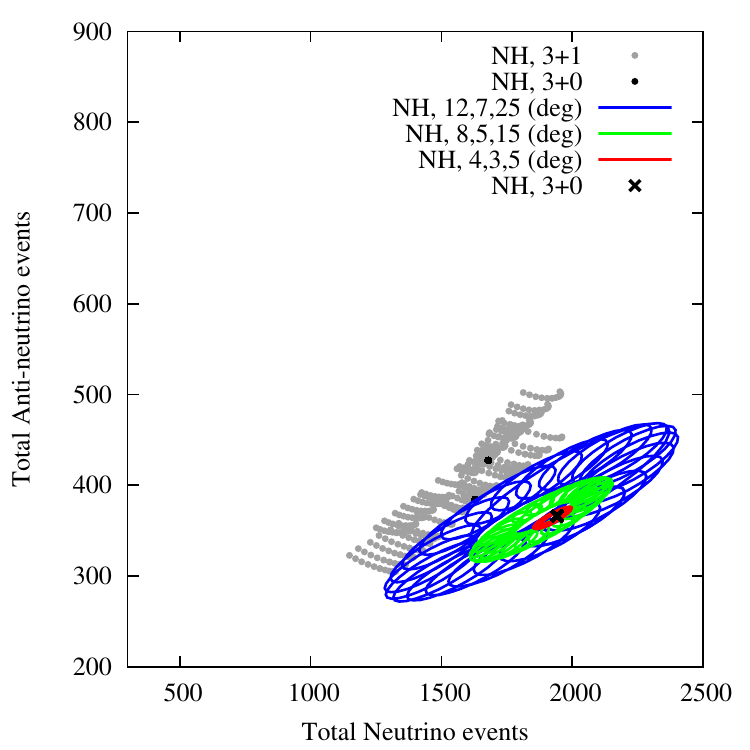}
\includegraphics[width=0.35\textwidth]{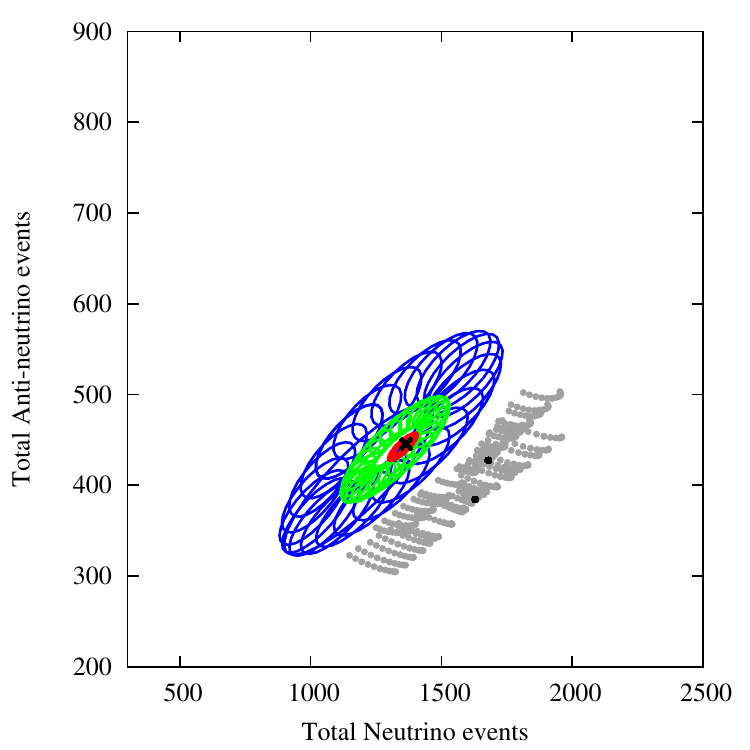}
\caption{\footnotesize{Neutrino and Anti-neutrino event rates for the 
DUNE experiment. The top panels are to explain the observed hierarchy
sensitivity while the bottom panels are for CPV. The event rates in blue,
green and red correspond to the simulated data with NH and the active-sterile
mixings as depicted in the keys. The black crosses correspond to true event rates
for NH and $\dcp=-90^\circ~(90^\circ)$ assuming 3+0 for the left (right) panel. The event rates in grey in the top panel correspond to the fit events 
rates assuming 3+1 where all the active-sterile oscillation parameters have been varied. The event rates in grey in the bottom panel correspond to 
fit event rates assuming 3+1 where the active-sterile mixing angles have
been varied and each of the three CP phases is either $0$ or $\pi$. In the 
top (bottom) panels, the assumed hierarchy for fit events is IH (NH). In 
the top panels, the black ellipses correspond to fit events with only
$\dcp$ varied assuming 3+0. In the bottom panels, the black dots 
correspond to fit events with $\dcp = 0, \pi$ assuming 3+0.}}
\label{eventsplots}
\end{figure}

This subsection attempts to obtain  a better understanding of the changes that occur in CP and hierarchy sensitivities at long baselines (as seen in subsections \ref{CPviolation} and \ref{MassHierarchy}) in the presence of a sterile neutrino. It provides, in effect, an alternative way of looking at the qualitative behaviour  exhibited by the sensitivities in Figs. \ref{CPnovat2k} through \ref{MHdune} and complements the discussion in the subsections above. We use DUNE as our primary example, but the features discussed below are, in general, displayed in the figures corresponding to other experiments as well, \eg ~those for NOvA + T2K or for HK.

In Fig. \ref{eventsplots}, we try to explain the results that we obtained 
for the hierarchy and CP violation sensitivities of the DUNE experiment
using total (energy-integrated) neutrino and anti-neutrino event rates. The top panels of Fig. \ref{eventsplots} refer to  the hierarchy sensitivity behavior
while the bottom panels are for CP violation. The regions shown in red, green, 
blue and the  black cross correspond to true oscillation parameters (or the data). 
The regions in grey, the black dots and the  black ellipses 
correspond to the test oscillation parameters (or the fit). 

The data events on the left plots are for true NH and true $\da=-90^\circ$ 
while the data events on the right plots are for true NH and true $\da=90^\circ$.
The grey fit events in the top panels correspond to test IH and all 3+1
($\ta, \tb, \tc, \db, \dc$ in addition to $\da$) 
parameters varied, while the black ellipse in the top panel corresponds
to test IH and variation of $\dcp$ only, assuming 3+0. The grey fit events in the bottom panel
correspond to test NH and test CP-conserving combinations of $\da, \db, \dc$, while the black
dots in the bottom panel correspond to test NH and test $\dcp=0,\pi$; assuming 3+0.
Broadly, it
can be said that the sensitivities will be better if the data events and the 
fit events are well separated (ignoring sensitivity coming from spectral information). The sensitivity patterns can be understood if we focus on the following two features:
\begin{itemize}
\item The fit event rates region increases drastically as we go from 3+0
to 3+1 (black ellipses and dots compared to grey regions). Note that we marginalise over the entire allowed 3+1 oscillation parameter space.
\item For small mixing angles $\theta_{i4}$, the event rate regions are
small compared to their size for  large mixing angles (red compared to green and blue).
\end{itemize}

Thus, we can make the following deductions which are reflected in the sensitivity plots.
\begin{itemize}
\item Compared to 3+0, the 3+1-small mixing angles reduce the sensitivity of
DUNE to hierarchy or CP violation because although the increase in the
data event regions is small, the fit event regions cover significantly more area, decreasing the separation between the two types of regions, and consequently, the sensitivity. This is visible in the green regions in the left panels of Figs. \ref{CPdune} and \ref{MHdune}.
\item As we go from small mixing angles to large mixing angles, the area covered by the  data
event rates increases significantly. Thus, there are oscillation parameters for which
the data events are either very far or very close to the fit event regions compared
to 3+0. Consequently, the 3+1 sensitivities can be much better or much worse
compared to the 3+0. This is demonstrated, for instance,  by the blue regions in the left panels of Fig. \ref{CPdune} (both better and worse; for CP) and Fig. \ref{MHdune} (mostly worse; for hierarchy).
\item Fig. \ref{eventsplots} also makes it  easier to understand the behavior with respect to true $\da$.
As we go from $\da=-90^\circ$ (left plots) to $\da=90^\circ$ (right plots), the 
fit rates approach data rates in the case of hierarchy and hence the sensitivity
decreases, as seen in the coloured regions of both the left and right panels of Fig. \ref{MHdune}. However, in the case of CP violation, the fit rates are symmetrically
placed around the data region and therefore, the sensitivity remains roughly
the same in both quadrants, as is manifested in  the left and right panels of Fig. \ref{CPdune}.
\end{itemize} 

\subsection{To what extent can DUNE ignore the presence of the 3+1 sector if it is present?}
\begin{figure}[t]
\centering
\includegraphics[width=0.3\textwidth]{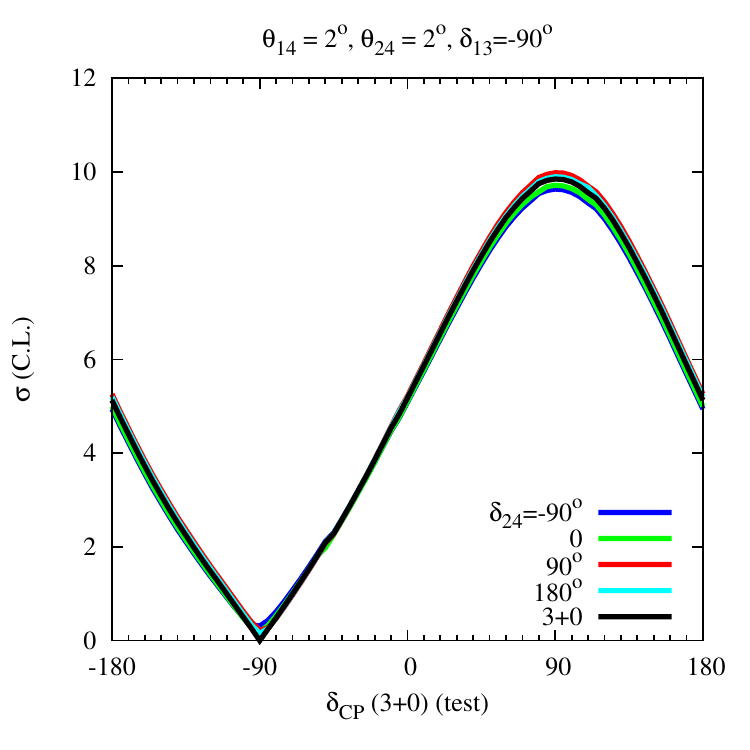}
\includegraphics[width=0.3\textwidth]{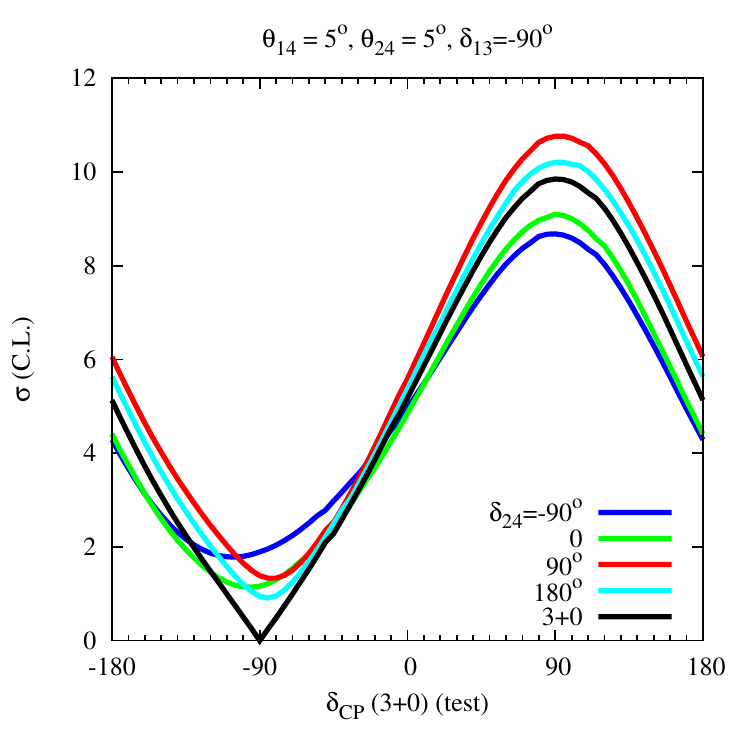}
\includegraphics[width=0.3\textwidth]{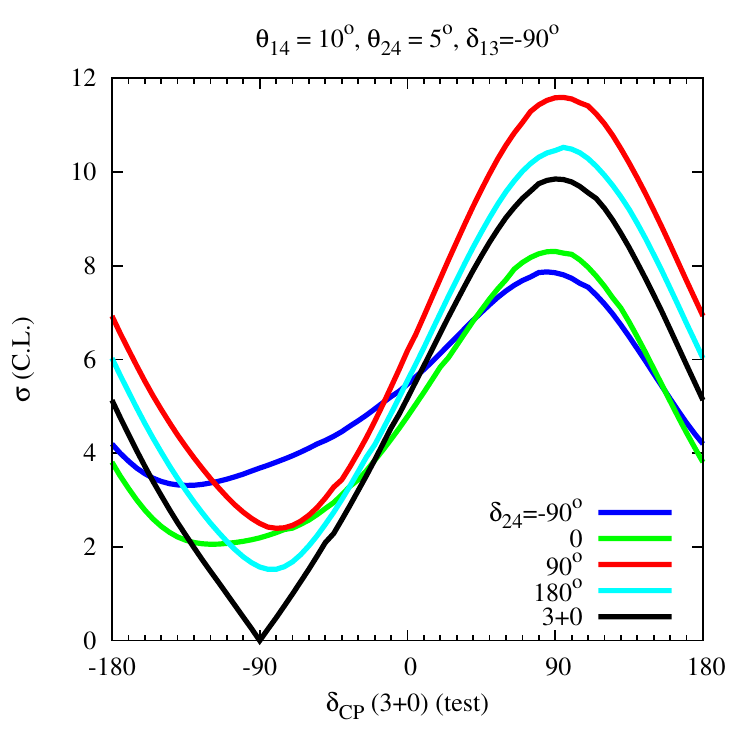}
\caption{\footnotesize{$\sigma = \sqrt{\Delta \chi^2}$ as a function of
test $\dcp$ for the DUNE experiment. We contrast 3+1 as the 
true case with 3+0 as the test case. Results are shown for 
different 3+1 oscillation parameters. These plots are 
for true $\da=-90^\circ$. True $\db$ values have been varied 
as shown in the key. The left, middle and right panels correspond to true 
$\ta = \tb = 2^\circ$, $\ta = \tb = 5^\circ$ and $\ta = 10^\circ, \tb = 5^\circ$ ($\ie$$\sin^22\theta_{\mu e}$ equal to 
$6\times10^{-6}$, $2\times10^{-4}$ and $9\times10^{-4}$)
respectively. We assume $\tc=\dc=0$.
Both true as well as test hierarchy is assumed to be normal.
 The black curve shows the usual sensitivity when both true and
test are 3+0.}}
\label{CPbias-90}
\end{figure}

In this subsection we explore the  question  whether
the CP-related measurements at DUNE can get affected by assuming
the absence of sterile neutrinos 
when they exist in reality, but are obscure because of very 
small mixings. To explore this point in greater detail, we carry
out the following exercise. We assume that the true scenario
is 3+1 with sufficiently small active-sterile mixings such that
$\sin^2 2\theta_{\mu e} \leq 10^{-3}$, $\ie$ below the currently stated sensitivity of short-baseline experiments planned or underway at Fermilab \cite{Antonello:2015lea, Zennamo, Gamez, Camilleri}. For such small values
of $\ta$ and $\tb$, $\tc$ is rendered inconsequential in $\pme$;
therefore, we assume $\tc=\dc=0$ for simplicity. We choose several 
true values of $\da$ and $\db \in [-180^\circ, 180^\circ]$.
In the fit, we assume 3+0 and vary $\dcp\in[-180^\circ,180^\circ]$,
calculating the sensitivity at each test $\dcp$. We fix the 
remaining  3+0 oscillation parameters in the fit at their true values.
The  results are shown in Fig. \ref{CPbias-90}, and the  details of our assumptions and chosen values are as described in the figure caption.

It is possible to obtain the following information from the curves in Fig. \ref{CPbias-90}:
\begin{itemize}
\item They show the  allowed test $\dcp$ region. For example,  in the 
3+0 scenario, DUNE should be able to measure $\dcp$ with a 
precision of $\pm60^\circ$ at $3\sigma$ C.L. if true $\dcp=-90^\circ$.
\item They demonstrate the  sensitivity at which the experiment can exclude CP 
conservation i.e. test $\dcp = 0, 180^\circ$ if the  true $\dcp=-90^\circ$.
\item Finally, they help answer the question: Can 3+0 be excluded if 3+1 is true?
If for all test $\dcp$
values, $\dxx > \rm{N}^2$, then we can say that 3+0 can be excluded
at $\rm{N}\sigma$.
\end{itemize}
We observe in Fig. \ref{CPbias-90} that if $\ta$ and $\tb$ are as small as 
$2^\circ$ each (left panel), then the results obtained are essentially the same even 
if the wrong theoretical framework of 3+0 is considered while fitting the data.
For angles as large as $5^\circ$ (middle panel), there are visible but  not highly significant deviations from the 3+0 results. Depending on the true value of $\db$, the precision
with which DUNE measures $\da$ can improve or worsen by $\sim10^\circ$.
However, the predicted best fit is around the right test $\dcp$ ($=-90^\circ$) 
in most  cases. An offset, of $\sim 15^\circ$, is obtained depending
on the true value of $\db$. For the 3+0 case, DUNE can exclude CP conserving
test $\dcp=-180^\circ$ at $5\sigma$. This sensitivity, in the 3+1 case, can vary from 
$4$-$6\sigma$. For $\ta=\tb=5^\circ$, DUNE can exclude 3+0 at $1$-$2\sigma$ depending on the true value of $\db$. For still larger values
of mixing angles like $\ta=10^\circ, \tb=5^\circ$, we observe similar features
except that the $\dxx$ are larger i.e. DUNE is more likely to exclude 3+0. 

\begin{figure}[t]
\centering
\includegraphics[width=0.3\textwidth]{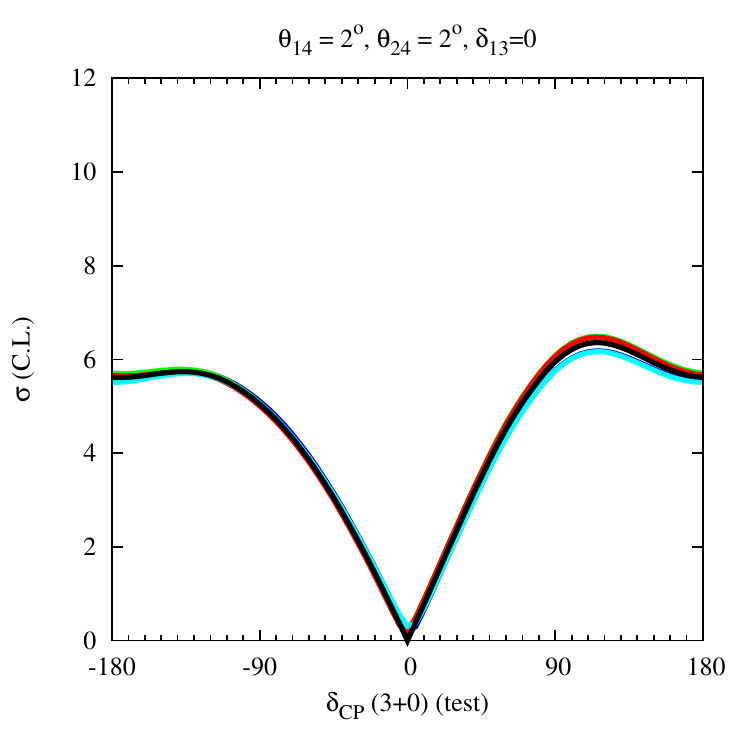}
\includegraphics[width=0.3\textwidth]{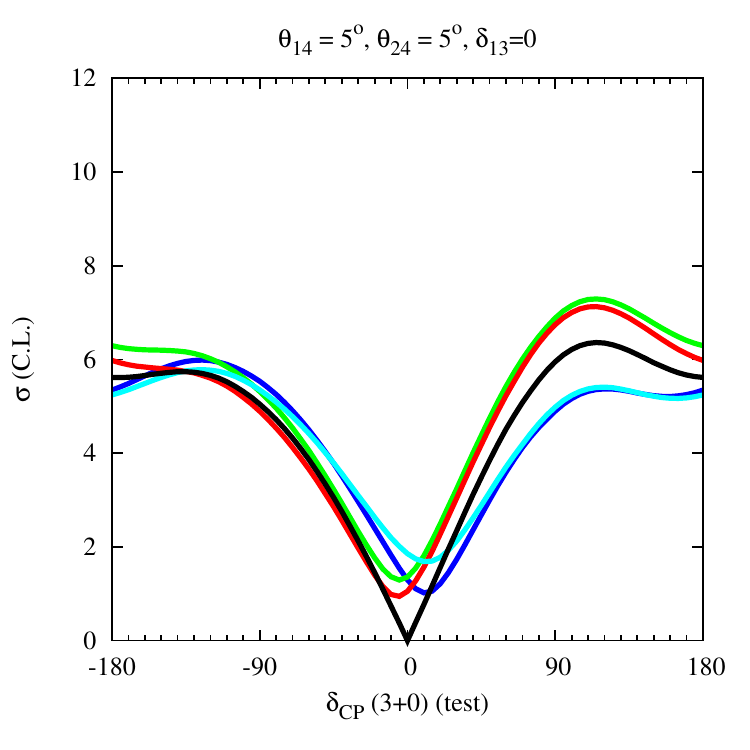}
\includegraphics[width=0.3\textwidth]{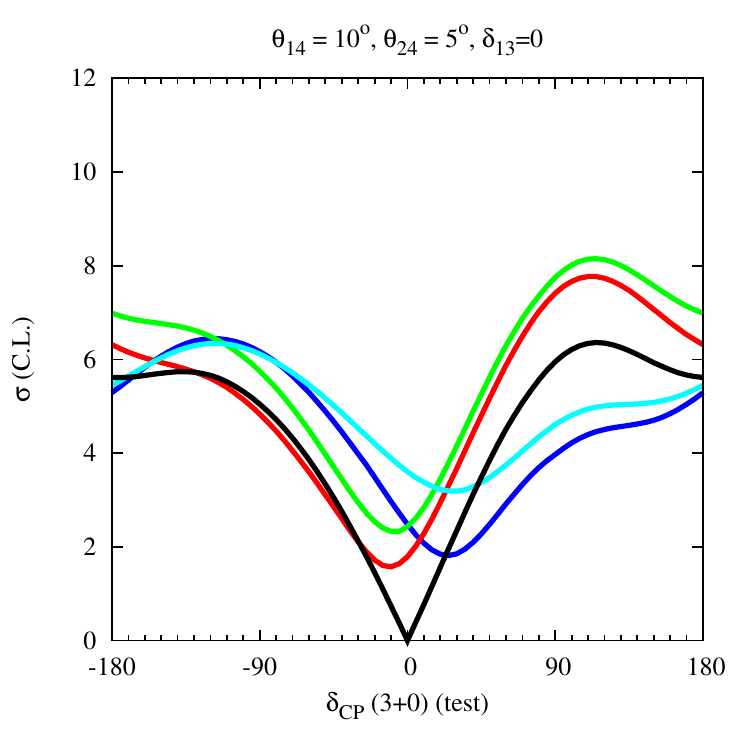}
\includegraphics[width=0.3\textwidth]{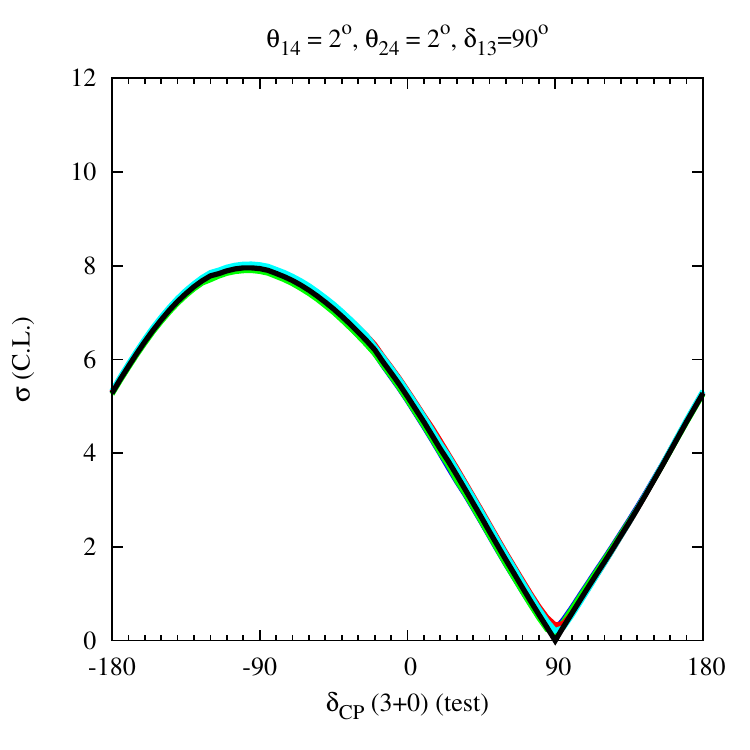}
\includegraphics[width=0.3\textwidth]{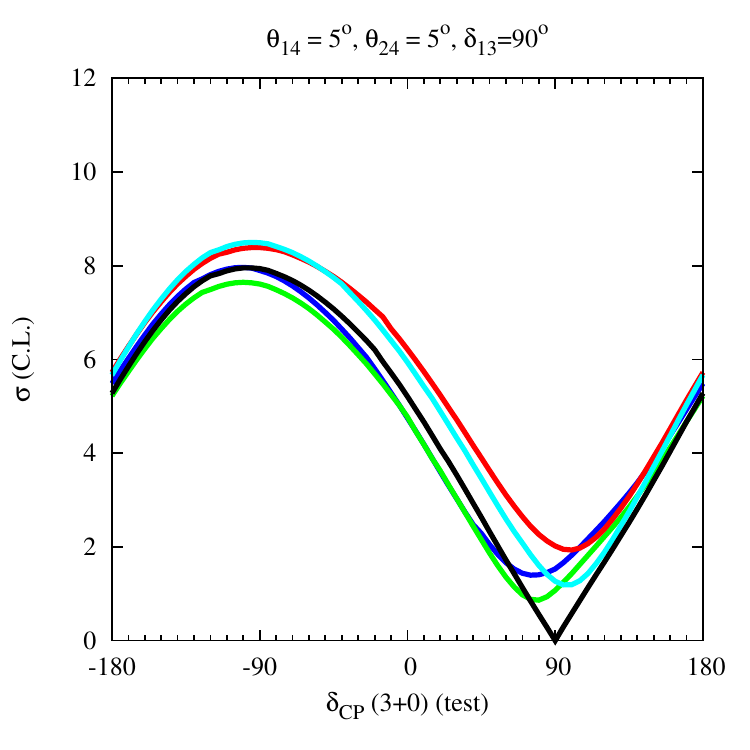}
\includegraphics[width=0.3\textwidth]{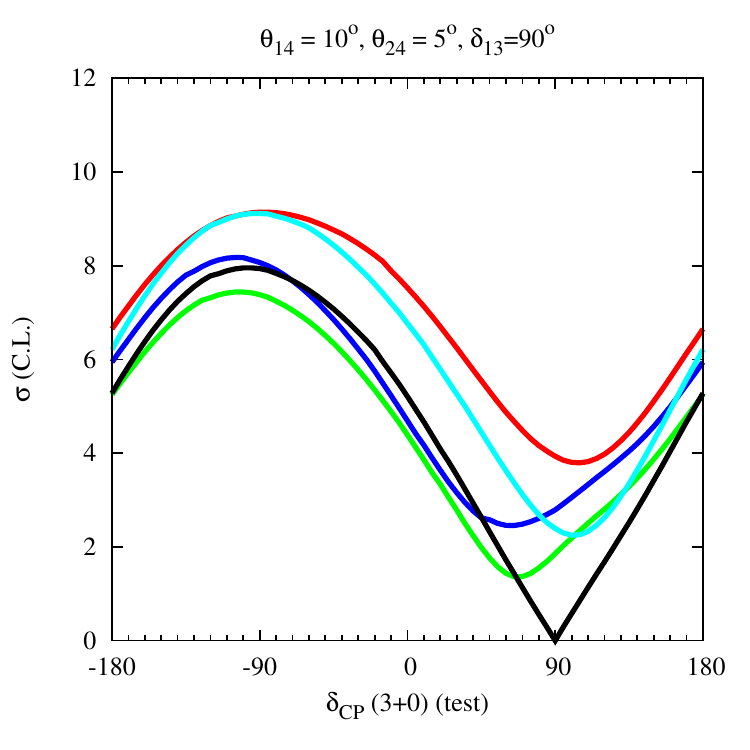}
\caption{\footnotesize{Same as Fig. \ref{CPbias-90}, but for $\da=0$
(top panel) and $\da=90^\circ$ (bottom panel).}}
\label{CPbias-multi}
\end{figure}

This exercise has been repeated  for true $\da = 0~\rm{and}~90^\circ$  in Fig. \ref{CPbias-multi}, with similar results.

\subsection{Determining the $\delta$ phase responsible for CP violation in 3+1}
\label{CPsource}
Suppose that the  results of short-baseline experiments indicate the presence of a $\sim$1 eV neutrino, and, in addition, DUNE finds evidence of CP violation. 
Assuming this situation, we attempt  to examine the extent and accuracy  with which DUNE can identify the source of this violation, $\ie$  the particular  3+1 phase ($\ie$ $\da$, $\db$ or $\dc$ ) associated with it. To simplify matters, we assume  near-maximal-allowed values 
of the sterile mixing angles; i.e. $(\ta, \tb, \tc): (12^\circ, 7^\circ, 
25^\circ)$ and also  fix the true value of $\dc$ to 0, in order to better bring out 
the effect of the other two CP phases.

\begin{figure}[H]
\centering
\includegraphics[width=0.49\textwidth]{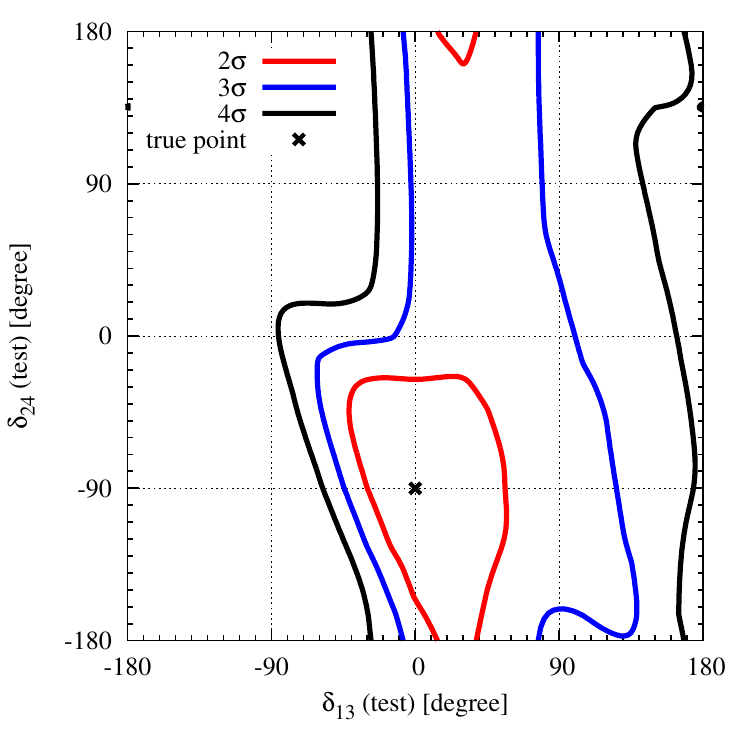}
\includegraphics[width=0.49\textwidth]{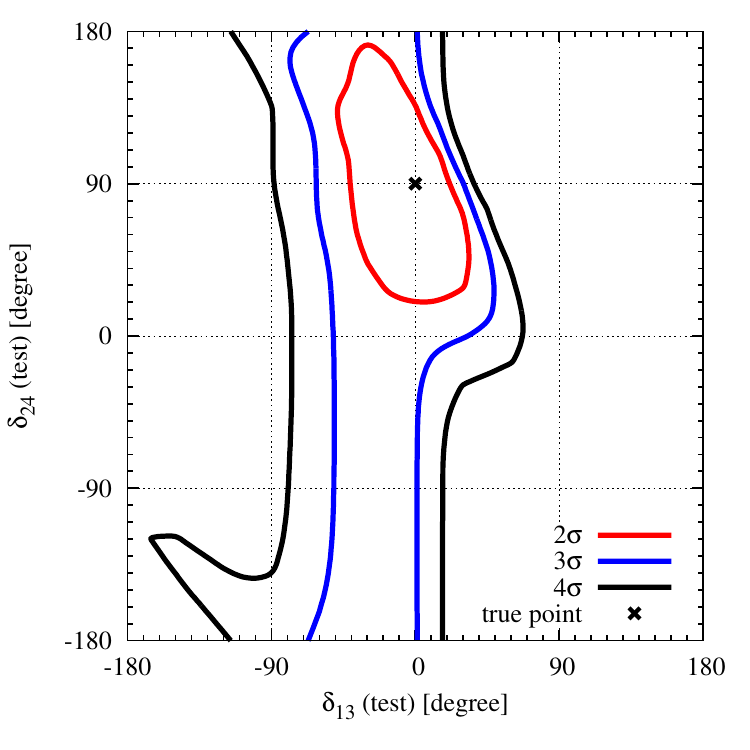}
\includegraphics[width=0.49\textwidth]{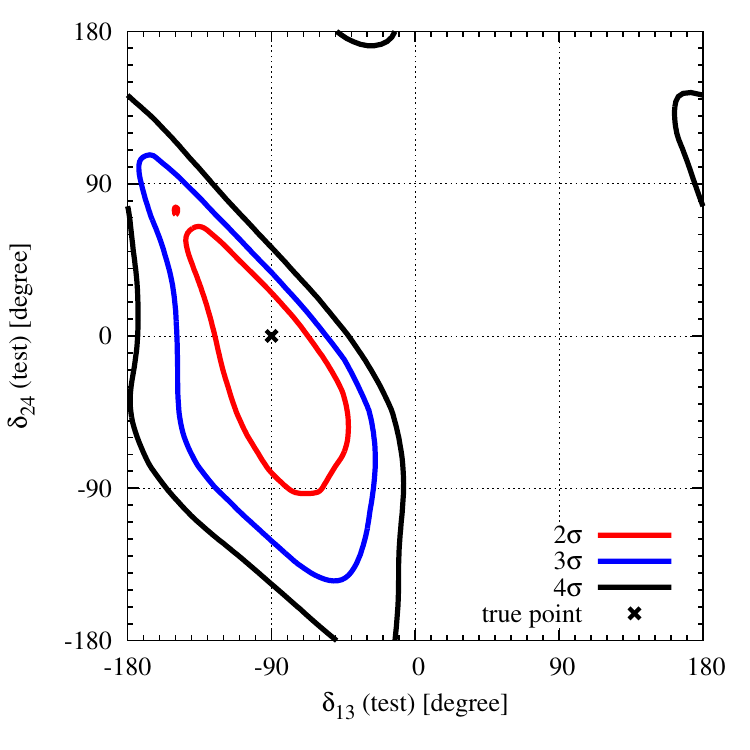}
\includegraphics[width=0.49\textwidth]{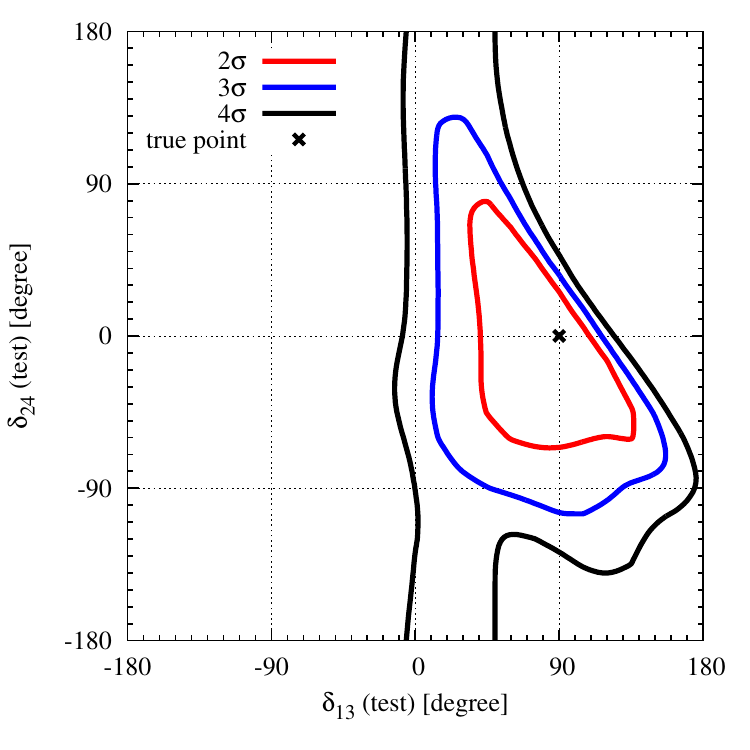}
\caption{\footnotesize{The allowed-region contours in the test
$\da$ - test $\db$ plane for DUNE. True $(\ta,\tb,\tc)$ were
taken to be $(12^\circ, 7^\circ, 25^\circ)$ and true $\dc=0$.
Results have been shown for true $(\da,\db) = (0,\pm90^\circ)$
and $(\pm90^\circ, 0)$. $2\sigma$, $3\sigma$, and $4\sigma$
results (corresponding to $\dxxmin = 6.18$, $11.83$ and $19.33$
respectively for two degrees of freedom) have been shown.
The true point in each plot has been shown with a cross.}}
\label{CPdetermination}
\end{figure}

 Results are shown for the 
following four combinations of true $(\da, \db)$: $(0, \pm90^\circ)$ 
and $(\pm90^\circ, 0)$. These correspond 
to the situation where one of the phases is maximally CP violating, 
while the other is CP conserving. In the fit, we marginalised over the
test active-sterile mixing angles and the test CP phase $\dc$. The 
$2\sigma$, $3\sigma$ and $4\sigma$ allowed regions in the 
test $\da$ - test $\db$ plane are shown in Fig. \ref{CPdetermination} 
for the above-mentioned four true combinations (these plots can also
be used to find out the precision with which DUNE can measure the 
CP phases; see \cite{Berryman:2015nua, Agarwalla:2016xxa} for
such results).

The top panels of Fig. \ref{CPdetermination} correspond to the
choice of true $(\da,\db)$ being $(0, -90^\circ)$ (left) and $(0, 90^\circ)$
(right). In the left panel, we see that test values of $(\da,\db)$ close to 
$(90^\circ, 0)$ and $(90^\circ, -180^\circ)$ are allowed within $3\sigma$ 
contours. Thus, it might be concluded that attributing CP violation 
unambiguously to $\db$ may not be possible at $3\sigma$. 
 In the right
panel, we see that test $(\da,\db)$ close to $(-90^\circ, 0)$
and $(-90^\circ, 180^\circ)$ are allowed, but at $4\sigma$. Similar
conclusions can also be drawn for the plots in the bottom panel
where distinguishing maximally CP-violating $\da$ and CP conserving
$\db$ from maximally CP-violating $\db$ and CP conserving $\da$
may not be possible at $4\sigma$.

\subsection{How large do active-sterile mixings need to be before DUNE becomes sensitive to their presence?}

\begin{figure}[t]
\centering
\includegraphics[width=0.45\textwidth]{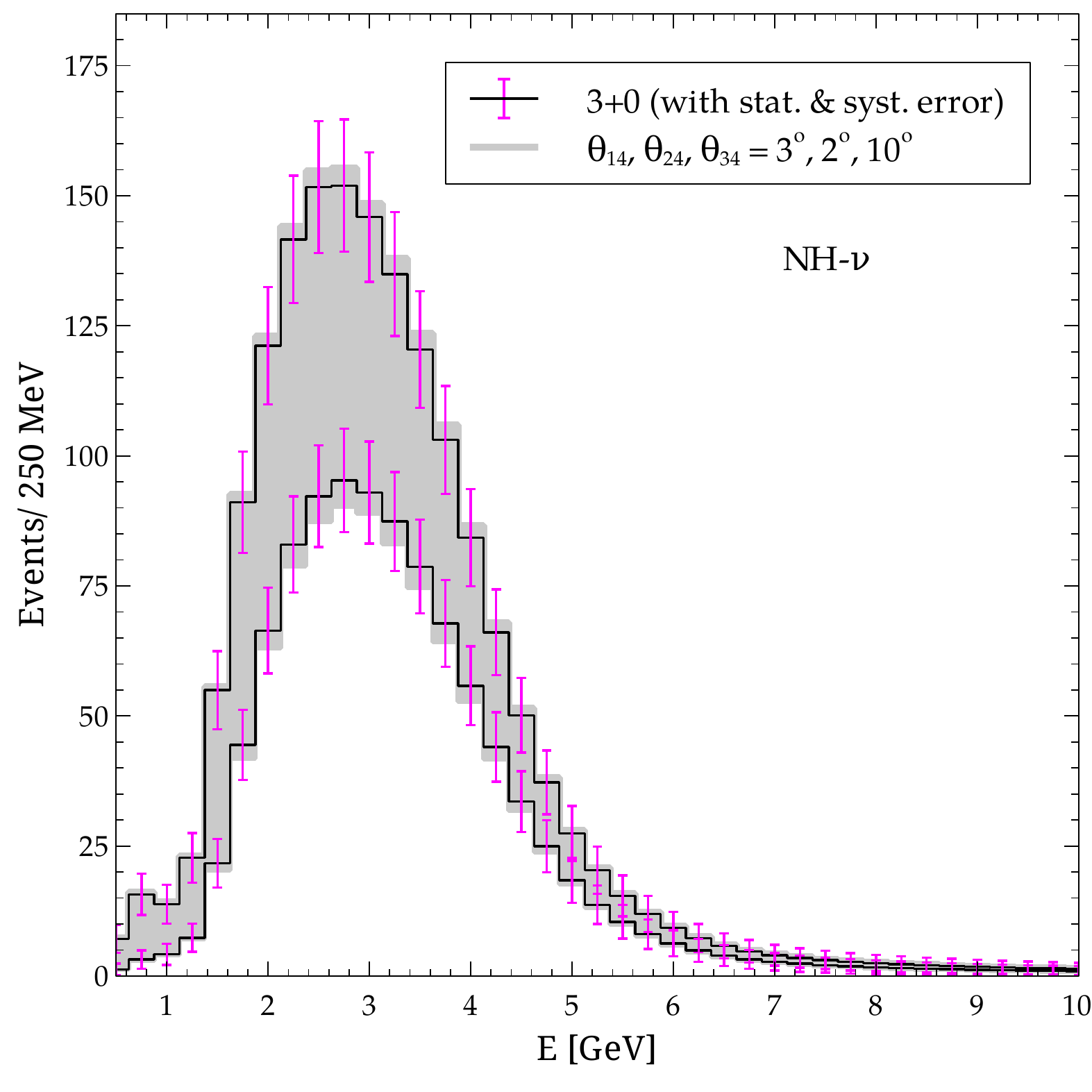}
\includegraphics[width=0.45\textwidth]{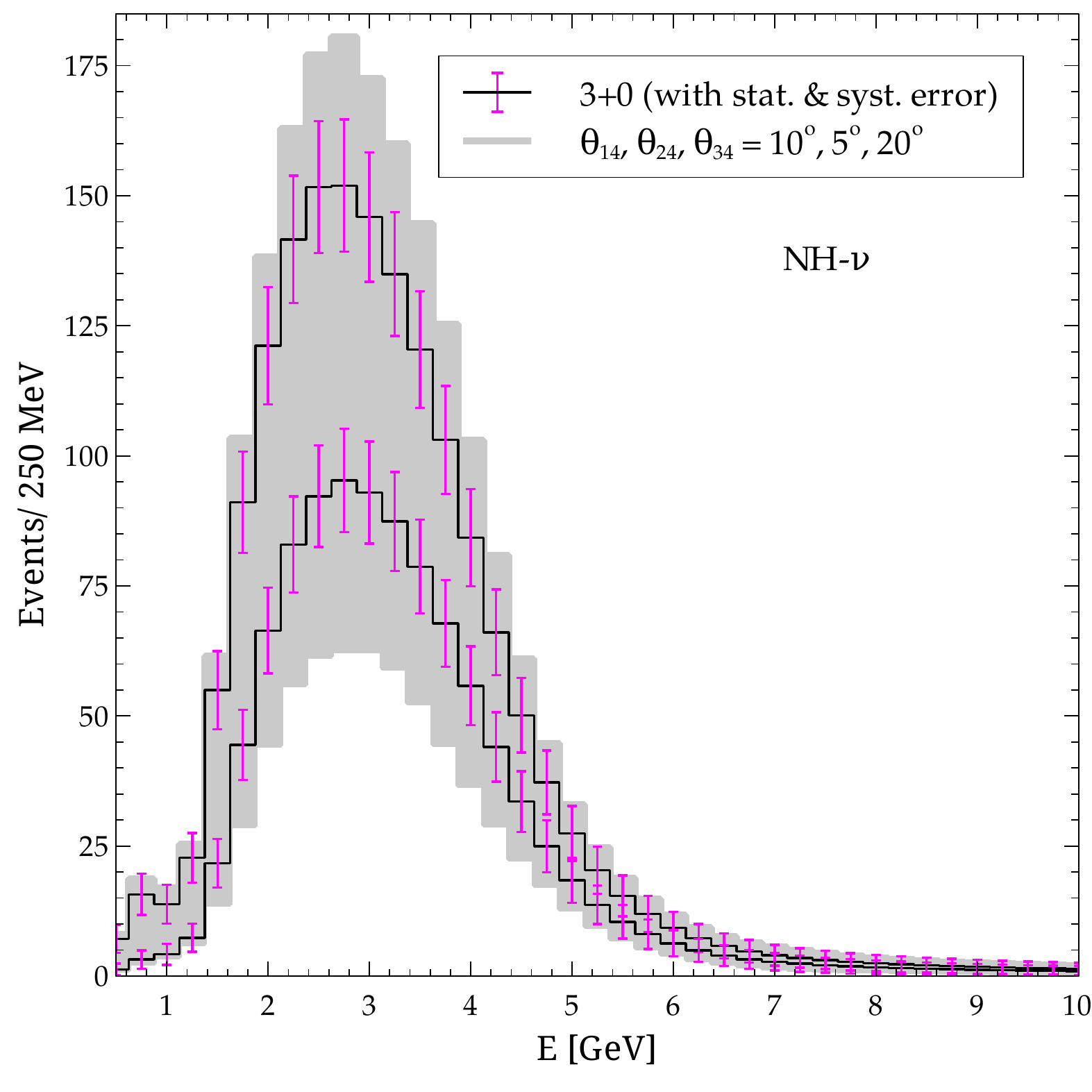}
\caption{\footnotesize{Neutrino event rates for the DUNE 
experiment as a function of the reconstructed neutrino 
energy. The black lines show the maximum and the 
minimum event rates corresponding to a variation of 
$\dcp$ in 3+0. Also shown are the corresponding statistical ($\sqrt{\text{Event no.}}$) and systematic error ($2\%$)\cite{elizabeth} added in quadrature for each energy bin. The grey band corresponds 
to the maximum and minimum event rates assuming 3+1 with 
$\ta, \tb, \tc = 3^\circ, 2^\circ, 10^\circ$  (left panel), 
$\ta, \tb, \tc = 10^\circ, 5^\circ, 20^\circ$  (right panel) and $\da, \db, \dc$
varied in $[-180^\circ, 180^\circ]$. Only the channel $\numu 
\rightarrow \nue$ has been considered with 5 years of 
$\nu$-running assuming normal hierarchy.}}
\label{DUNEeventserror}
\end{figure}

The Short Baseline Neutrino (SBN) program at Fermilab aims to conclusively establish
the existence or else to place stringent constraints on the possible existence of the sterile neutrinos. At short baselines,
the $\pme$ oscillation probability is sensitive only to the mass-squared
difference $\delta m^2_{41}$ and an effective mixing angle given by
$\sin^2 2\theta_{\mu e} = \sin^22\ta\sin^2\tb$. For $\delta m^2_{41}
\sim 1~\textrm{eV}^2$ induced oscillations, the SBN program 
can exclude at  $3\sigma$ only $\sin^2 2\theta_{\mu e} \geq 0.001$ 
\cite{Camilleri}. It is natural to ask how tightly active-sterile mixings need to be excluded to ensure that DUNE measurements can be safely interpreted without taking the possible existence of sterile neutrinos into account. Phrasing this question another way, we ask whether active-sterile mixings corresponding to $\sin^2 2\theta_{\mu e} < 0.001$ can be detected by the DUNE far detector. Fig. \ref{DUNEeventserror} (left panel)
throws some light on this question. Here, we have compared the CP bands
in event rates assuming 3+1 and 3+0 for very small mixing angles - 
$\ta, \tb, \tc = 3^\circ, 2^\circ, 10^\circ$ ($\sin^2 2\theta_{\mu e} 
\approx 0.00001$). We see that the 3+1 band is completely degenerate
with the 3+0 band (incorporating the errorbars\footnote{The error for each energy bin is the quadrature sum of the statistical ($\sqrt{\text{Event no.}}$) and the systematic error ($2\%$)\cite{elizabeth}. The value is an estimated expected value assuming the presence of a highly capable near-detector\cite{elizabeth}.}).
Thus, for
such small mixing angles, it seems that neither the Short Baseline
experiments nor the DUNE experiment may see evidence of new 
physics attributable to sterile neutrinos. In the right panel of the same figure, however, we have chosen $\ta, \tb, \tc = 10^\circ, 5^\circ, 20^\circ$ ($\sin^2 2\theta_{\mu e} 
\approx 0.0009$). These larger values, which correspond to an effective mixing angle at  the $3\sigma$ sensitivity of the short-baseline experiments, lead to enhanced effects. The grey band now extends significantly beyond the expected event rates for 3+0, even after accounting for errors. This provides a suggestive estimate  of how large  the mixing angles need to be before sterile neutrino effects at DUNE start to be discernable. 

We stress that short-baseline experiments are  significantly more sensitive, by design,  to CP conserving  oscillatory effects  in 3+1 compared to long-baseline experiments, and hence remain the definitive test by which presence or absence of sterile neutrinos can be established. Long-baseline experiments, however, can become sensitive to the presence of this sector if CPV is present. Moreover, matter enhances the effects of CP, further enabling these experiments to become, in a sense, complementary detectors of sterile neutrinos \cite{Gandhi:2015xza}.  Signals that could possibly be interpreted as those for sterile neutrinos  at long-baseline experiments like DUNE would, however,  remain  supportive rather than definitive evidence, because they could concievably  be mimicked by other physics beyond the SM (see, for instance, \cite{Masud:2015xva, Coloma:2015kiu, deGouvea:2015ndi, Forero:2016cmb, Liao:2016hsa, Huitu:2016bmb, Bakhti:2016prn, Masud:2016bvp, Rashed:2016rda, Coloma:2016gei, deGouvea:2016pom, Masud:2016gcl, Blennow:2016etl} for the effect of propagation NSI on long baselines, especially at DUNE in the context of CP Violation and Mass hierarchy.).

%====================================================================================
\section{ Conclusions and Summary}
\label{summary}

This work examines how sensitivities of long-baseline experiments to the MH and CP violation are affected and altered in the presence of a sterile neutrino. It attempts to quantitatively examine questions raised in  \cite{Gandhi:2015xza}. While we use  DUNE as our benchmark example, we study these sensitivities for T2K, HK, and  NOvA also. Depending on the values of sterile mixing angles and phases, the sensitivities can be both significantly enhanced or suppressed compared to the 3+0 case. We examine and discuss the reasons for this behaviour using the total event rate as a tool.

We also examine the ability of DUNE to pinpoint the origin of CPV, if such violation is detected by it. We find that while the discovery potential for the violation is large, determining its origin ($\ie$  ascribing it unambiguously to either the 3+0 phase $\dcp$ or one of the other  phases $\db, \dc$, present in 3+1) is much more challenging. Indeed, $3\sigma$
 determination of the phase (or phases)  responsible for CPV could prove very elusive both if sterile neutrinos are shown to  exist, or if their existence cannot be conclusively ruled out by the short-baseline experiments. If the latter is the case, 
 we ask how tightly one must then bound the sterile-active mixing angles to ensure that DUNE data can be safely interpreted without taking the possible existence of sterile neutrinos into account. In the process, we find that DUNE may exhibit signals hinting at  the presence of a sterile sector even if the relevant mixing angles lie below the sensitivity of the planned short-baseline experiments. However, the ability of long-baseline efforts like DUNE to signal the presence of this sector, while highly valuable, must remain complementary to an essential and primary short-baseline thrust aimed at discovering evidence of short-wavelength oscillations with convincing redundancy. As emphasized both in  \cite{Gandhi:2015xza} and this work, the sensitivity of DUNE to sterile neutrinos has qualitatively different origins and  depends on interference effects and the matter enhancement  of the corresponding mixing angles and CP violating phases. These are certainly aspects which do not lie within the physics ambit of short-baseline experiments; however, the resulting signals could also be mimicked by other new physics, preventing their  unambiguous interpretation if considered in isolation.

 In summary, DUNE provides  strong and valuable complementarity in the search for a sterile sector to be conducted by the short-baseline experiments. It would be useful, in our view, to incorporate this capability into the thinking and efforts currently underway towards optimizing its design.

%====================================================================================

%====================================================================================
\begin{acknowledgments}
We gratefully acknowledge many helpful discussions on DUNE fluxes and errors with  Mary Bishai and Elizabeth Worcester. We are also thankful to Dan Cherdack, Georgia Karagiorgi, Bryce Littlejohn, Poonam Mehta and Lisa Whitehead for very useful discussions. We also acknowledge the use of the HPC cluster facility at HRI for carrying out the numerical computations used in this work. RG acknowledges support and hospitality from the Neutrino and Theory Divisions at Fermilab while this work was in progress. DD, RG and  MM acknowledge support from the DAE Neutrino project at HRI. Fermilab is operated by Fermi Research Alliance, LLC under contract no. DE-AC02-07CH11359 with the US Department of Energy.
\end{acknowledgments}

%====================================================================================
\bibliographystyle{apsrev}
\bibliography{references}
\end{document}